\documentclass[12pt, letterpaper]{article}
\pdfoutput=1

\usepackage[utf8]{inputenc}
\usepackage{graphicx,epsfig,changepage}
\usepackage{makeidx,setspace,epstopdf}
\usepackage[svgnames]{xcolor}
\usepackage{subcaption}
\usepackage{amsmath,amssymb,amsfonts,amsthm,amsbsy}
\usepackage{mathrsfs,bold-extra}
\usepackage[shortlabels]{enumitem}

\usepackage[square,numbers,merge,comma,sort&compress]{natbib}

\usepackage[blocks]{authblk}

\makeatletter
\normalsize
\makeatother

\usepackage[top=3cm, bottom=3cm, left=3.75cm, right=3.75cm,heightrounded,marginparwidth=1.5cm, marginparsep=1cm]{geometry}

\usepackage[nottoc]{tocbibind}

\usepackage{hyperref}
\hypersetup{colorlinks=true,citecolor=blue,linktoc=page,filecolor=black,linkcolor=blue,urlcolor=Green,
pdfstartview={FitH}}

\title{%
\vspace{-2em}
\centering\boldmath\Large\bfseries%
Interaction between macroscopic quantum systems and gravity
\vspace{1em}
}

\author{Antonio Gallerati\,\footnote{antonio.gallerati@polito.it}}
\affil{\makebox[1em][c]{Politecnico di Torino, Dipartimento DISAT, corso Duca degli Abruzzi 24, 10129 Torino, Italy}}

\author{Giovanni Modanese\,\footnote{giovanni.modanese@unibz.it}}
\affil{Free University of Bozen-Bolzano, Faculty of Science and Technology, I-39100 Bolzano, Italy}

\author{Giovanni Alberto Ummarino\,\footnote{giovanni.ummarino@polito.it}}
\affil{\makebox[1em][c]{Politecnico di Torino, Dipartimento DISAT, corso Duca degli Abruzzi 24, 10129 Torino, Italy}}
\affil{\makebox[1em][c]{National Research Nuclear University MEPhI, Kashirskoe hwy 31, 115409 Moscow, Russian Federation}}

\date{}
\begin{document}

\maketitle

\begin{abstract}
\noindent
We review experiments and theoretical models about the possible mutual interplay between the gravitational field and materials in the superconducting state or other macroscopic quantum states. More generally, we focus on the possibility for quantum macrosystems in a coherent state to produce local alterations of the gravitational field in which they are immersed. This fully interdisciplinary research field has witnessed a conspicuous progress in the last decades, with hundreds of published papers, and yet several questions are still completely open.
\end{abstract}

\maketitle

\newpage

\begingroup
\onehalfspacing
\tableofcontents
\endgroup

\pagebreak

\section{Introduction}
It was in the 1960's that physicists began to wonder if it was possible to use superconductors as sensitive gravity probes, especially for detecting new effects predicted by General Relativity, like gravitational waves and gravitomagnetic fields \cite{DeWitt:1966yi,papini1966london,papini1967detection}.

In the 1970's, S.\ Hawking proved that quantum effects in classical spacetime had an important back-reaction on spacetime itself, leading to the evaporation of black holes. At the time, the quantum theory of gravitation was still in its infancy, but in the following 50 years it has grown into a very active research field (maybe hyperactive), with essential applications at least to early cosmology and to scattering processes at very high energy \cite{amelino2013quantum}.

The progress in superconductivity has been even more impressive. High-T$_\text{c}$ superconductors comprise today a variety of materials and microscopic structures. Josephson junctions, like other mesoscopic systems, have become important elements in quantum computing (\cite{choi2001coherent} and refs.); arrays of intrinsic Josephson junctions in high-T$_\text{c}$ materials work as synchronized THz emitters \cite{rajasekaran2016parametric}, etc.\ -- just to mention a few practices well beyond the initial field of application of superconductors.

Over the years it has become clear that the traditional geometro-dynamical scheme employing classical Einstein field equations plus a matter tensor $T_{\mu \nu}$ might be inaccurate when matter is in a macroscopic quantum state. Since the mechanism of superconductivity in high-T$_\text{c}$ superconductors cannot be described by the BCS theory, the use of an effective wave equation (the time-dependent Ginzburg-Landau equation) is very important in this context (see Sect.\ \ref{explGL}). On the other hand, quantum properties of gravitation could also play a role in interactions between gravity and superconductors (Sect.\ \ref{role}).

Are there for these interactions any observational consequences of the quantum nature of matter, or of gravity, or both? Can we obtain some information from existing experiments? In this review we will actually start by collecting the experimental evidence (Sect.\ \ref{experim}), then after a brief treatment of gravitational antennas (Sect.\ \ref{antennas}), we will review theoretical models (Sect.\ \ref{theory}).

The only pre-existing review on this subject is the article by Kiefer and Weber from 2005 \cite{Kiefer:2004hv}, which in the present work is updated and extended in a substantial way, especially concerning the experimental part. Another related review (actually a volume with thematic chapters) was published in 2012 \cite{modanese2012gravitybook}; besides obviously missing all the later developments, it does not cover the theoretical subjects treated in Sect.\
\ref{antennas} of this review and only a minor part of those in Sect.\ \ref{theory}.

The macroscopic quantum systems addressed in this work are, as most frequently understood, superconductors and superfluids. Nevertheless, general theoretical techniques involving the gravity-induced quantum phase (Sect.\ \ref{phase}) and the Ginzburg-Landau formalism (Sect.\ \ref{explGL}) can be applied also to other macroscopic quantum systems, like e.g.\ Bose-Einstein condensates.

\section{Experimental results}
\label{experim}

The late 1980's had seen the publication in the mainstream literature of new claims of non-conventional gravitational effects, possibly requiring modifications of standard theoretical approaches based on General Relativity. On one side there were measurements of gravitational forces in the sub-millimeter range, leading to the so-called ``fifth force'' hypothesis (see \cite{fischbach1992six} for an early review and \cite{lee2020new} for very recent measurements and an update on the current status). On the other side observations were reported of an anomalous weight reduction of rotating gyroscopes \cite{hayasaka1989anomalous}.

These claims, plus the idea of a possible amplification role of macroscopic coherence in superconductors, probably contributed to motivate the work of Podkletnov and, later, of Tajmar. Actually the specific results of \cite{hayasaka1989anomalous} were soon contradicted by other experiments and later a new wealth of data about rotating gyroscopes were obtained, e.g., in the Probe-B experiment \cite{everitt2011gravity}. It is still true today, however, that the Newton gravitational constant is known with lower precision than any other fundamental constant. This might partly be due to some dependence of the force upon setups and materials used in the measurements, a dependence that is not accounted for by General Relativity. See about this, e.g. \cite{westphal2021measurement} and refs., and the review \cite{gillies1997newtonian}; the latter also covers early claims of possible weak gravitational screening effects due to interposed bodies (compare recent work in \cite{unnikrishnan2001anomalous} and \cite{coisson2002new}).

\subsection{The 1990's: the first experiment by Podkletnov and weight anomalies at the superconducting transition}

The experiments by E. Podkletnov with YBCO superconductors have considerably modified the scenario of the search for gravity-superconductors interactions, even though his results have been, until now, neither confirmed nor disproved. These experiments involve a complex setup and their theoretical modelling is also inevitably complex, thus substantial research investments would be required to exit from the present stalemate.

YBCO (Yttrium Barium Copper Oxigen) is a high-T$_\text{c}$ superconductor with a critical temperature of about 93 K, which can be easily manufactured in films or small sintered disks and operated at liquid nitrogen temperature. In Podkletnov's experiments, however, large disks with more complex crystal structure were used (up to 27 cm diameter) and cooled below 70 K using liquid helium. By ``first experiment'' we mean the one with rotating and levitating disks, in the two versions of 1992 and 1997. By ``second experiment'' we mean the experiment of 2002 with high-voltage discharges through a superconducting electrode (Sect.\  \ref{ep2}).

In the first version, published in 1992 \cite{podkletnov1992possibility} a large YBCO disk (with 145 mm diameter and 6 mm thick) was magnetically levitated using a single coil operated at frequencies of 50-100 Hz. The disk was maintained at temperatures below 60 K in the vapours of liquid helium. It was a sintered single-layer disk with relatively small grains. In most trials the disk was also irradiated with a radio-frequency field and spun to several thousands rpm using additional magnets.

The original purpose of the experiment was to test the material under those conditions. It was noticed, however, that small test masses suspended above the disk were apparently losing a small fraction of their weight, up to 0.05\%. The measurement procedure was criticised by some authors \cite{de1995alternative,unnikrishnan1996does}, who dismissed the reported effect as due to systematic errors.

In 1997 Podkletnov published an improved version. After being accepted by J.\ Phys.\ D, the article was withdrawn due to a controversy between the author and the university of Tampere, and subsequently posted in the Los Alamos physics database \cite{podkletnov1997weak}. The new experiment featured a larger YBCO disk in a ring configuration (27 cm outer diameter, 8 cm inside diameter and 1 cm thick). The material had a bi-layer structure as described below. The disk levitated inside a steel cryostat thanks to several coils operated in a two-phase mode to drive it up to 5000 rpm.

The material had been prepared using a ceramic sintering technique followed by partial melt-texturing, in which the upper surface was heated to 1200 $^0$C using a planar high-frequency inductor. As a consequence, the upper part of the disk (6-7 mm thick) had an ortho-rhombic structure typical of the quench and melt growth process, with $T_\text{c}=94\pm 2 $ K and $J_\text{c} \simeq 15000$ A/cm$^2$, while the transition temperature of the lower layer was $60 \pm 10$ K. The operation temperature was slightly below 70 K. The best results were obtained under an applied radio-frequency of the order of 3.2 -- 3.8 MHz.

Samples of different materials were hung over the cryostat and weighed using sensitive balances. They all appeared to lose the same fraction of their weight, typically 0.3 to 0.5\%. The external boundary of the ``shielding cylinder'' was quite clear (no more than 2 cm). A detailed analysis of possible systematic errors is given in the paper, ruling out several spurious effects.

\subsubsection{Replication attempts}

(For a more detailed historical description see \cite{modanese2012gravity,hathaway2009gravitational}.)

In a joint effort between the University of Alabama in Huntsville and NASA Marshall Space Flight Center, several single-layer superconducting disks were fabricated in 1999 \cite{koczor1999fabrication}; shortly thereafter, Podkletnov was hired as a consultant and a bi-layer disk was fabricated. As concerns measurements of gravity effects, however, only static tests on a commercially purchased 4-inch single-layer disk were published, with limited results \cite{li1997static}.

In 2001, Woods et al.\ reported an unsuccessful replication attempt with disks of diameter up to 3 inch \cite{woods2001gravity}. The authors argued that many of the conditions for a gravity effect had not been fulfilled, including conditions on the disk material structure and the levitating and rotating system.
It should be mentioned that no visual evidence (pictures or video) of the original experiments exists, and this has represented an obstacle for the replication attempts.

The replication attempt by Hathaway et al.\ \cite{hathaway2003gravity} came closer to Podkletnov's setup, still some major differences remained; in particular, the disk was sustained and rotated only mechanically, at lower velocity and lower temperature. No weight modifications or gravity-like forces were detected down to the 0.001\% level. According to a discussion by Robertson in \cite{modanese2012gravity}, the lack of a proper power input into the superconductor could have been key for the missing effect.

Also contained in \cite{modanese2012gravity} is a phenomenological analysis of the first Podkletnov experiment, discussing its incompatibility with classical General Relativity, energetic balance, properties of the field configuration associated to the reported cylindrical ``partial shielding region'', relations to the equivalence principle and to the London field in rotating superconductors.

\subsubsection{Transient weight modifications at the superconducting transition}

In 1996 the American engineer and inventor J.\ Schnurer reported claims of transient anomalous gravitational effects caused by small melt-textured YBCO disks \cite{modanese2001possible}. The disks had a typical diameter of 2 cm and a thickness of 5 mm. They were cooled down below T$_\text{c}$ and then pinned in levitation above a permanent magnet, at room temperature. After a while the disks returned normal and fell down slowly on the magnet, the fall lasting approximately 1 to 3 seconds. At that moment, small target glass rods hung on the vertical of the disk (with an insulating plate in between) exhibited temporary weight changes of the order of 1\%. The weight was measured with an electronic balance connected to the target by a rod and a counterweight. Approximately 10\% of the total 400 trials gave positive results, clearly distinguished from the noise, signalling that some of the parameters triggering the effects were not under control.

A more accurate replication was done in 1998-99 in a master thesis at the Polytechnic of Turin, Italy \cite{binello1999studio}. The observed weight changes were smaller, of about $10^{-4}g \simeq 10^{-3}$ m/s$^2$, but measurements were more reliable, because the target was the plastic plate of a precision analytical balance encased in a glass box and placed on an anti-vibration table, 1 meter above the superconductor. The weight modification quoted above was estimated from the oscillations of the balance plate assuming that the perturbing force lasted 1 s; if the actual duration was shorter, then the inferred strength should increase in inverse proportion.

In 1997 H. Reiss made accurate measurements of the weight of sintered YBCO pellets contained in a plastic capsule while the sample passed through its critical temperature \cite{reiss1999possible,reiss2003weight}. Such experiments are difficult because the whole sample is first cooled by immersion in liquid nitrogen, and then the evaporation of nitrogen residues causes a slow change of weight, while the supposed anomalous effect is evidenced by some sudden spikes around this background. Measurements performed with this technique clearly differ from those of Schnurer because their aim is to detect possible weight changes of YBCO itself, not of separated suspended targets. Later Tajmar et al.\ published null results of related measurements performed with liquid nitrogen \cite{tajmar2005weight} and without liquid nitrogen \cite{tajmar2009measuring}; these results have been in turn examined by Reiss and Hathaway in further work \cite{reiss2006minimum,hathaway2012experimental}.

\subsection{The 2000's: the second Podkletnov experiment and those by Chiao, Tajmar, Poher}
\label{ep2}

In 1997-2002 E. Podkletnov built in Moscow a low pressure gas chamber in which high voltage discharges occurred between and YBCO cathode and a copper anode, at temperatures between 50 and 70 K \cite{podkletnov2003investigation}. The discharge circuit comprised a Marx generator with a capacitor array and was able to produce voltages up to 2000 kV and peak currents of 10 kA. The discharge chamber had a diameter of 1 m and length of 1.5 m. The cathode (diameter 80-120 mm, thickness 7-15 mm) consisted of two layers. The internal layer (the one not facing the anode) was not superconducting at the operation temperature, because a rare-earth element was included into YBCO. Both layers were fabricated by sintering using seeded oxygen-controlled melt-texture growth. A large solenoid allowed the application of a magnetic field to the chamber, and a smaller coil could be used to trap a field in the cathode.

The shape of electric discharges at low pressure (less than 1 Pa) depended on the voltage, changing from a single direct spark to a more diffused discharge at voltage above 500 kV, lasting less than $10^{-5}$ s. In coincidence with such ``flat'' discharges, the emission of an anomalous non-e.m.\ radiation was observed, which propagated in a focused beam along the axis of the chamber, towards the anode and beyond it, reaching large distances without attenuation. From the impulse transferred by this radiation to small targets it could be determined that the energy $E$ carried by the beam was of the order of 1 mJ, but its momentum was much larger than $E/c$ (the momentum carried by e.m.\ or gravitational waves). In fact, it looked more like a short-lasting force beam, suggesting a comparison with the small (but longer-lasting) lifting force generated by the YBCO disks of the previous experiment.

\subsubsection{Replication attempts and further developments}

A replication attempt with a 50\% scaling in size was conducted by T.\ Junker starting in 2004 at the Inst.\ f\"ur Gravitationsforschung (Germany) \cite{junker2012setup}. The technical challenges encountered were immense, and the theoretical modelling also very difficult \cite{junker2008conditions}. The results were inconclusive. For a brief description see \cite{modanese2012gravity}. A phenomenological analysis is given in \cite{robertson2012quantum}.

Another scaled-down replication attempt was reported by Tajmar and Lorincz \cite{lorincz2015design}. In fact, their device is more similar to that of C.\ Poher (see \ref{further}), since the discharge is not in vacuum and the maximum voltage is 2.2 kV. In an improved follow-up work of 2016 \cite{lorincz2016null} the same authors came to the conclusion that with good e.m.\ shielding the accelerometer effects can be attributed to acoustic noise.

In 2012 Podkletnov reported measurements of the effect of the anomalous radiation beam on a laser beam crossing it with an angle of 0.1 degrees, giving an interaction region approx.\  57 m long \cite{podkletnov2012study}. The intensity of the laser spot was measured to decrease by 7-10\% during the discharge. An attempt was also described of measuring the propagation speed of the anomalous radiation using two piezoelectric sensors connected to synchronized atomic clocks and placed along the radiation beam at a distance of 1211 m.

\subsubsection{The experiment by R.\ Chiao et al}
R.Y.\ Chiao conducted in the early 1980's, also in collaboration with J.\ Anandan, studies on the possibility of constructing antennas for gravitational radiation using superconductors and superfluids \cite{anandan1981gravitational,anandan1982gravitational,chiao1982interference}. In 2003 he published with W.J.\ Fitelson and A.D.\ Speliotopoulos the results of a new experiment aiming to test the efficiency of conversion of e.m.\ waves into gravitational waves and viceversa, mediated by YBCO as a possible quantum transducer. Theoretically, the concept was based on a new minimal coupling rule for the coupling of electron spin to curved spacetime in general relativity, mediated by a quantum fluid. (This coupling has been discussed in detail by Kiefer and Weber in their review \cite{Kiefer:2004hv}, with the conclusion that it is ``questionable''.)

The experiment was performed at the Univ.\ of California in Berkeley and was not conclusive, setting an upper limit of $1.6\cdot 10^{-5}$ on the conversion efficiency of YBCO ar liquid nitrogen temperature. The work is carefully described and discussed in an arXiv preprint \cite{chiao2003search}. This was however not published in a journal. A related paper appeared in a Conference Proceedings in 2006 \cite{chiao2006macroscopic}. For related proposals by Chiao et al.\ see also \cite{chiao2004conceptual,chiao2006interface,Minter:2009fx,inan2017interaction}.

More in detail, the idea of the experiment was to use a YBCO sample to convert microwave e.m.\ radiation into gravitational radiation and then use a second sample to back-convert its far-field component into e.m.\ radiation. The two samples were contained in two separated closed metallic containers, whose e.m.\ coupling was measured to be extremely small. In order to prevent transitions out of the macroscopically coherent superconducting quantum state, the frequency of the microwaves was chosen to be below the gap frequency of YBCO. The conservation of angular momentum in the conversion process was satisfied by first converting the e.m.\ field into a quadrupolar far-field radiation pattern with angular momentum 2.

In the analysis of the experiment, the authors discussed if there may exist quantum many-body enhancements to the classical conversion efficiency. A further question was, how does a broken-symmetry ground state like that of a superconductor interact with a dynamically changing spacetime, such as that associated with a gravitational wave; such question has never been explored experimentally. Finally, the authors conclude that since one of the conditions for a good coupling of a quantum antenna and transducer to the gravitational wave sector is extremely low dissipative losses, the choice of YBCO as the material medium for an Hertz-type experiment may have not been good, given its D-wave character.

\subsubsection{Other works on high-frequency gravitational waves}
It seems appropriate to cite here other papers concerning the possible generation through superconductors of high-frequency gravitational waves (called HFGWs in the literature; see \cite{dehnen1981generation,stephenson2003application,li2007detection,baker2008analyses}).

The works by R.C.\ Woods \cite{woods2005manipulation,woods2006enhanced,woods2007exploitation} do not describe experiments, but contain sophisticated phenomenological and feasibility studies.

Articles by G.\ Fontana look at possible quantum gravitational emission in pair tunnelling between type-I and type-II superconductors \cite{fontana2004design,fontana2012high}.

Landry and Paranjape have analyzed the possible features of a graviton laser \cite{landry2016graviton}.

\subsubsection{The experiments by Tajmar and collaborators}
\label{tajmar}

The main experiments by Tajmar et al.\ were performed in the period 2006-2011 \cite{tajmar2007measurement,tajmar2008investigation,tajmar2009anomalous,tajmar2010fiber} using detectors placed near rotating superconductors maintained at various temperatures. Standard ``spring'' accelerometers were used, along with fiber optical gyroscopes, which measure rotating gravitational fields affecting the phase of light beams traversing closed paths.
The idea behind the experiments and their interpretation was that rotation generates gravitomagnetic and gravitoelectric fields obeying equations of the Maxwell-Einstein type, but including large amplification factors of the sources \cite{tajmar2006gravitomagnetic}.

According to the detailed review of Lewis \cite{lewis2012tajmar}, another motivation for the experiments was the discrepancy found in the tests of superconducting gyroscopes for the Gravity Probe-B experiment (see also the theoretical model of Hauser and Dr\"oscher cited below). Tajmar and De Matos \cite{tajmar2006gravitomagnetic} hypothesized that such discrepancy might be explained by a very large increase of the gravitomagnetic field in a rotating superconductor. According to standard General Relativity, their observed gravitomagnetic field would be equivalent to one produced by a white dwarf star!

Independent measurements performed by Graham et al.\ in 2008 \cite{graham2008experiment} with a lead disk and one ring-laser gyroscope gave results declared within $1 \sigma$ from a null result. For the results of the Tajmar group the figure would be $11 \sigma$.

In agreement with the Tajmar group (working in Austria), also the Graham group (in new Zealand) reported a phenomenon of ``parity violation'', namely that gravitomagnetic fields produced by the disk have different strengths, depending on the rotation sense. For a detailed discussion see \cite{lewis2012tajmar}.

The rotating Nb ring employed by Tajmar in 2007 \cite{tajmar2007measurement} had a radius of 69 mm, thickness 6 mm, height 15 mm. Its angular acceleration could reach 1000 rad/s$^2$, and the deceleration 5500 rad/s$^2$. Depending on the various conditions of acceleration and temperature, the fields detected by the accelerometer had strenghts in the range from $10^{-4}$ to $10^{-3}$ m/s$^2$.

In the experiments of 2008 \cite{tajmar2008investigation} Tajmar used a set of 4 gyroscopes mounted in 3 positions. Signals were observed from Al, Nb and YBCO disks, but only for clockwise rotation. The experiments of 2009 \cite{tajmar2009anomalous} showed that liquid and superfluid helium rotating with the superconductors had an effect on the results. Finally, in 2011 Tajmar et al.\ \cite{tajmar2011evaluation} re-interpreted their results, attributing the anomalous effects to some kind of acoustic noise. The reason was that new experiments with modified equipment measured gyroscopes signals about 2 orders of magnitude lower compared to earlier results.

Hauser and Dr\"oscher maintain that in their theoretical description of Tajmar's experiments these different results can be explained by different thermal environments in which the detectors operated. They have proposed a new geometric approach for describing physical interactions, called Extended Heim Theory \cite{droscher2008gravity,hauser2010emerging,hauser2017gravity,hauser2022gravity}. This approach predicts the existence of three gravitational fields (both attractive and repulsive), of stable neutral leptons and of particles of imaginary mass, which might be a component of dark matter. According to \cite{droscher2008gravity}, anomalous gravitomagnetic fields similar to those reported by Tajmar et al.\ might have been generated in orbit during the Gravity Probe-B experiment and might be partially responsible for the observed gyroscope misalignment.

\subsubsection{Further claims of anomalous gravitational effects}
\label{further}

In this Section we mention other peer-review experimental works describing effects that can be interpreted as due to anomalous gravitational interactions.

The experiments by C.\ Poher \cite{poher2011physical,poher2017enhanced} display some similarities with the second Podkletnov experiment, because they involve high voltage discharges through YBCO electrodes. The voltage is much lower than in Podkletnov's discharge chamber; actually there is no gas chamber, and the discharge is triggered by a solid-state switch. The discharge currents are of the order of several kA and cause a violent recoil of YBCO electrodes (but not of dummy metal electrodes used as benchmark, and also not for YBCO above T$_\text{c}$). The discharges also appear to generate a force or radiation beam with high energy/momentum ratio, similar to that observed by Podkletnov in his second experiment, but the beam is weaker and diverging. The force has been measured with accelerometers. Similarities and differences between the experiments have been discussed in detail in \cite{modanese2013comparison}, whose author has personally witnessed many demonstration of Poher's device in France, while this was not possible with Podkletnov's experiment, hosted in a restricted area in Moscow.

The experiment by Godin and Roshchin \cite{roshchin2000experimental} is a careful reconstruction and adaptation of a Searl machine with rotating magnets. It does not employ any superconductors. Possible connections with the first Podkletnov experiment and the Tajmar experiments arise from the presence of rotating materials containing strong trapped magnetic fields, which in turn produce high-frequency induced electric fields. A number of anomalous effects were observed, including a reduction in the weight of the rotating multi-magnets (core plus ``rollers''). This effect may be connected to results obtained by another group using an accurate accelerometer connected to falling magnets \cite{porcelli2022new}. Weak unexpected changes in the local gravity acceleration were observed, indicating the possible influence of a generalized quantum entanglement phenomenon.

\section{Gravitational antennas}
\label{antennas}

The direct observation of quantum gravity effects in condensed matter, involving gravitons dynamical interactions with other quantum fields at the microscopic level, is in general a very difficult task.
A different approach could originate from the study of unconventional, macroscopic states of matter. In this regard, one should consider quantum macrosystems existing in nature, like superconductors and superfluids. The latter can be thought as large systems featuring a macroscopical coherent phase, suitably described by order parameters. It could be then possible to formulate (and observe) a possible interplay between the extended, coherent system and the surrounding gravitational field \cite{DeWitt:1966yi,papini1966london,papini1967detection,hirakawa1975superconductors,
Ciubotariu:1991zw,anandan1994relgra,podkletnov1992possibility,Modanese:1995tx,
Modanese:1996zm,agop1996gravitational,li1991effects,ahmedov1999general,agop2000some,
Modanese:2001wv,Wu:2003aq,hathaway2003gravity,Kiefer:2004hv,Quach:2015qwa,Ummarino:2017bvz,
Ummarino:2019cvw,Ummarino:2020loo,Ummarino:2021vwc,Ummarino:2021tpz,atanasov2017geometric,
atanasov2018gravitation}.
The coupling with the current flow without resistance in superconductors was exploited to use the latter as a sensitive detection systems, in particular for gravitational waves \cite{Papini:1970cw,anandan1982gravitational,Cerdonio:1984zz,anandan1984relativistic,
anandan1985detection,carelli1985coupling,mashhoon1989detection,Preparata:1989gd,Peng:1990ca,
Peng:1990cc,Peng:1991ua,peng1991electrodynamics,li2007detection,Minter:2009fx,
inan2017interaction,inan2017new,Hammad:2020xta}.\par
The first studies about gravitational (super)conducting detectors considered the possibility of converting gravitational in electromagnetic radiation \cite{Papini:1970cw}. When acted upon by a gravitational wave, the material would emit electromagnetic radiation of the same frequency of the incoming wave, thus behaving as a device transforming part of the gravitons into photons.
%A gravitational detector based on this effect would then consist in a metal rod in the normal or superconducting state, together with the radio apparatus necessary to receive and amplify the induced electromagnetic waves.
The efficiency of this kind of detector depends on the ratio of the power absorbed by the receiver to the radiated gravitational power per unit area, averaged over a sphere.

\subsection{Resonant detectors}
A simpler proposal for the detection of gravitational waves, in normal and superconducting devices, relies in the analysis of the associated vibrations in the detecting system.\par
A normal antenna can be considered as a standard resonant structure where positive ions and electrons are bound and vibrate together under the influence of the wave.
If we denote with $\omega_0$ the bar natural frequency of vibration, we obtain a damping time
\begin{equation}
\tau_0 \gg \frac{1}{\omega_0}\:.
\end{equation}
A resonant eigenmode of the detector can be represented by a damped simple harmonic oscillator and the device response can be in general obtained from the equations of motion of the various mass elements of the antenna, moving under the influence of the gravitational wave, the elastic force and possible dissipative forces \cite{Misner:1974qy}
\begin{equation}
\frac{d^2x}{dt^2}+\frac{1}{\tau_0}\:\frac{dx}{dt}+\omega_0\,x^2=a\,e^{-i\,\omega\,t}\:,
\end{equation}
where $x(t)$ represent the distance from the center of mass, and $a\,e^{-i\,\omega\,t}$ is the gravitational driving acceleration that results from projecting the tidal force due to the gravitational wave onto the antenna. It is then possible to obtain the displacements as
\begin{equation}
x(t)= -a\,G(\omega)\,e^{-i\,\omega\,t}\:,
\end{equation}
with
\begin{equation}
G(\omega)^{-1}=\omega^2-\omega_0^2+i\,\frac{\omega\,\omega_0}{Q}\:,
\end{equation}
in terms of the response function $Q=\omega_0\,\tau_0$.\par\medskip
Let us now study superconductive materials. Superconducting devices are promising systems for the detection of gravitational waves because of their distinctive properties, namely zero resistance and perfect diamagnetism. The latter give rise to some important effects, like the presence of induced electric fields in the interior. Moreover, cooling the detection device below the critical temperature is also a good strategy in order to reduce one of the most important sources of noise, the random thermal fluctuations.\par
In a superconducting antenna, once the material is cooled below the critical temperature, negative charge carriers exhibit independent responses to the gravitational wave (no resistance is offered to superelectrons), while the ions still vibrate as damped harmonic oscillators.
We then find different oscillations modes for ions and superelectrons, determining the appearance of a net current (including both ion and London supercurrent) giving rise to a time-dependent magnetic potential $A(r,t)$.
%which is responsible for the creation of ion-supercurrent and London supercurrent, also decreasing the displacements of ions oscillations
The latter induces in turn an electric field $E(r,t)=-\tfrac{\partial A}{\partial t}$ which reacts on ions and superelectrons, decreasing the displacements of ions vibrations (especially in the $\omega\sim\omega_0$ case) and increasing those of superelectrons.
In addition, this time-dependent internal field plays an important role in ensuring that the Meissner effect is not violated: magnetic fields produced respectively by the net current induced by the gravitational wave and the one generated by $E(r,t)$ are cancelled out such that the total field in the interior vanishes, i.e.\ the vibrations of the magnetic field die away within the penetration depth. Also the ions contribute to cancel the time-dependent magnetic field: in the thin surface layer ions will vibrate differently from superelectrons, so that a screen current appears expelling the magnetic field from inside the antenna. Part of the energy of the gravitational wave will be transformed to electromagnetic energy, leading to a decrease in the vibrational energy.\par
Let us now assume that gravitational waves penetrate the antenna without attenuation and that the tiny vibrations of ions will not destroy superconductivity. According to a two-fluid model \cite{london1961superfluids}, we should consider the presence of positive ions, normal electrons and Cooper pairs.
We still represent the induced ions motion by damped simple harmonic oscillators, and their equations of motion in the detector's proper reference frame read
\begin{equation}
\frac{d^2x_\text{i}}{dt^2}+\frac{1}{\tau_0}\:\frac{dx_\text{i}}{dt}+\omega_0\,x_\text{i}^2
    =a\,e^{-i\,\omega\,t}+\frac{e}{m_\text{i}}\,E(x)\,e^{-i\,\omega\,t}\:,
\end{equation}
while the equations of motion of the Cooper pairs are written as
\begin{equation}
\frac{d^2 x_\text{e}}{dt^2}
    =a\,e^{-i\,\omega\,t}-\frac{e}{m_\text{e}}\,E(x)\,e^{-i\,\omega\,t}\:,
\end{equation}
$m_\text{i}$ and $x_\text{e}$ being the masses of the ion and electron, respectively. The displacements of ions and superelectrons then read
\begin{equation}
\begin{split}
x_\text{i}(t)&=\left(-a-\frac{e}{m_\text{i}}\,E(x)\right)\,G(\omega)\,e^{-i\,\omega\,t}\:,
\\[2ex]
x_\text{e}(t)&=\left(-a+\frac{e}{m_\text{e}}\,E(x)\right)\,\frac{1}{\omega^2}\,e^{-i\,\omega\,t}\:,
\end{split}
\end{equation}
the ions equation reducing to the one of a normal antenna in the absence of induced electric field. The ion-current and electron-current  are given by
\begin{equation}
\begin{split}
J_\text{i}&=2\,n\,e\,\frac{dx_\text{i}}{dt}=J_\text{i}^\text{(g)}+J_\text{i}^\text{(s)}
    =i\,2\,n\,e\,\omega\,G(\omega)\,a+J_\text{i}^\text{(s)}\:,
\\[2ex]
J_\text{e}&=2\,n\,e\,\frac{dx_\text{e}}{dt}=J_\text{e}^\text{(g)}+J_\text{e}^\text{(s)}
    =-i\,2\,n\,e\,\omega\,G(\omega)\,a+J_\text{e}^\text{(s)}\:,
\end{split}
\end{equation}
where the first contributions depend on the gravitational driving acceleration and where the ion and London electron supercurrents are given by
\begin{equation}
\begin{split}
J_\text{i}^\text{(s)}
    &=-\frac{2\,n\,e^2}{m_\text{e}}\:\frac{m_\text{e}\,\omega^2\,G(\omega)}{m_\text{i}}\:A(x,t)\:,
\\[2ex]
J_\text{e}^\text{(s)}&=-\frac{2\,n\,e^2}{m_\text{e}}\:A(x,t)\:,
\end{split}
\end{equation}
$n$ being the number of Cooper pairs per unit volume. The net current induced by the incoming gravitational wave turns then out to be a combination of electron and ion supercurrents, the latter determined by the presence of a varying magnetic vector potential. The total net current can be schematically expressed as
\begin{equation}
J=J_\text{e}+J_\text{i}+J_\text{n}\:,
\end{equation}
where $J_\text{n}$ is the normal current density.\par
%If we ignore the effects of the gravitational wave, the situation reduces to the London equations and there will be no net current, electric $E$ induced field and magnetic field in the interior of the superconductor.
As we already mentioned, the Meissner effect is preserved. In this regard, the total net current split in a surface screen current, which only exists in a very thin layer near the sample surface, and another contribution in the interior of the superconducting antenna. The same happens for the induced electric field, the consequent vibrations of ions and superelectrons adjusting such that there is no magnetic field inside the superconductor.\par
In \cite{Peng:1990ca,Peng:1990cc} Peng and Torr consider a Weber-type detector \cite{weber1984gravitons}, consisting in a superconducting bar antenna, a transducer that converts information about the antenna's vibrations into an electrical signal and a DC superconducting quantum interference device (SQUID) coupling system. For this kind of apparatus, we expect the electric field in the interior to decrease the ion-current and the displacements of ions, while increasing the electron-current and displacements of superelectrons \cite{Peng:1990ca}. In particular, it is possible to evaluate the differences between superconducting and normal antennas: if we consider the displacements of ions in the two cases, we find for a non-resonant situation ($\omega\neq\omega_0$)
\begin{equation}
x_\text{i}\:\simeq\:x\:,
\end{equation}
which implies a similar behaviour. If we focus instead on a resonant case ($\omega\simeq\omega_0$) the ratio between the ions displacements in the superconducting and normal device gives
\begin{equation}
\left|\frac{x_\text{i}}{x}\right|\:\simeq\:\frac{1}{1+\left(m_\text{e}\,Q/m_\text{i}\right)^2}\:,
\label{eq:ratioxix}
\end{equation}
implying that the behaviour of the superconducting device depends on how we choose the quality factor $Q$. For a normal antenna, the sensitivity is determined in terms of the ratio of the burst squared signal (depending on the mean change $|x|$) to the mean--square Gaussian noise (depending on the temperature) \cite{thorne1987gravitational}
\begin{equation}
R_\text{n}\:\propto\:\frac{|x|^2}{T_\text{n}}\:,
\label{eq:Rsensitivity}
\end{equation}
$T_\text{n}$ being the ``noise temperature'', determined by thermal noise in the antenna and in the sensors. The behaviour of latter parameter can be approximated as
\begin{equation}
T_\text{n}\:\propto\:\sqrt{T_\text{a}/Q}\:,
\label{eq:Tnoise}
\end{equation}
where $T_\text{a}$ is the temperature of the antenna. Therefore, to minimize thermal noise, the possibilities are a reduction in the system temperature or an increase of the quality factor $Q$.\par
If the antenna is cooled below the critical temperature $T_\text{c}$, the normal antenna becomes a superconducting one with the different properties we have previously described. However, we expect the above relations for the sensitivity to still approximately hold true even for a superconducting device, $x$ being replaced by $x_\text{i}$. We then find
\begin{equation}
R_\text{s}\:\propto\:\frac{|x_\text{i}|^2}{T_\text{n}}
\qquad\Longrightarrow\qquad
R_\text{s}\,\simeq\,
    R_\text{n}\:\frac{1}{1+\left(m_\text{e}\,Q/m_\text{i}\right)^2}\:,
\end{equation}
having also used \eqref{eq:ratioxix}. We then obtain that, in the resonant case $\omega\simeq\omega_0$, a superconducting antenna with optimized quality factor $Q$ has a signal-to-noise ratio $R_\text{s}$ that is reduced with respect to the $R_\text{n}$ one of a normal antenna, with a correspondent decrease in the elements displacements. On the other hand, the distinctive properties of a superconducting material would determine separated oscillations for ions and superelectrons in response to a gravitational wave, giving rise to peculiar currents in the sample, also responsible for preserving the Meissner effect. In particular, the induced internal electric field can provide a higher sensitivity with respect to the usual mechanical vibration detection.

\subsection{Moving superconductors}
In order to provide a description of the electrodynamics of moving superconductors, Anandan proposed
the covariant equation \cite{anandan1984relativistic}
\begin{equation}
-\hbar\,\partial_\mu\phi+\frac{2\,e}{c}\,A_\mu=\frac{2\,\xi}{c}\,u_\mu\:,
\end{equation}
where $u_\mu$ is the four-velocity of the superconductor and the function $\xi$ is determined requiring agreement with the Josephson equation. The above expression
assumes that, in the interior of a superconductor, the current density is directly proportional
to the velocity of the superconductor itself.\par
In \cite{Peng:1991ua,peng1991electrodynamics} the authors tried to develop a phenomenologically--covariant formalism to investigate the interaction between an arbitrarily moving superconductor and surrounding electromagnetic, gravitational and gravito-Maxwell fields, also considering the effects deriving from the presence of gravitational waves.
Starting from a set of generalized Einstein-Maxwell equations, the authors study the electrodynamics of arbitrarily moving superconductor. They obtain that both the motion of ions and the presence of forces acting on superelectrons determine the dynamics of a moving superconductors. The forces acting on ions affect the expression for the ions velocity, which turns out to be case dependent. The formalism also predicts the presence of induced currents and fields in the interior of a vibrating superconductor, validating the principle for a gravitational wave antenna.

\subsection{Charge separation and dynamical Casimir effect}
\label{subsec:Casimir}
In \cite{Minter:2009fx,Quach:2015qwa} the authors consider the occurrence of a possible charge separation effect in the interior of a superconductor, as a result of the interaction with an incident gravitational wave. The situation is characterized describing the Cooper pair density in the superconductor by a Ginzburg-Landau theory, while the ionic lattice is modeled by harmonic oscillator sites coupled to the gravitational wave. This would give rise to a difference in motion of the negatively-charged Cooper pairs (condensed into a non-local quantum mechanical eigenstate) and the positively-charged lattice ions, locally responding to the gravitational wave (the ionic lattice thereby oscillating in its deformation accordingly). In particular, the response to the gravitational wave is far less for the Cooper pairs than for the lattice: this would determine the charge separation effect, which can in turn be used to detect the passage of the wave itself, and a non-negligible gravitational Casimir effect in vacuum could take place \cite{Quach:2015qwa}.
The charge separation also induces an electric field which opposes the supercurrents induced by the gravitational wave, the resulting Coulomb interaction acting as a strong restoring force making the superconductor extremely stiff to the gravitational radiation. Therefore, the incoming wave will be radiated back out, and the superconductor will act as a mirror to gravitational waves \cite{Minter:2009fx}, see also Section \ref{subsubsec:mirror}.\par
The described behavior can be associated to a quantum rigidity (associated with the BCS energy gap) exhibited by the Cooper pairs, which makes them relatively non-responsive to the gravitational wave. By contrast, the phonon modes of the ionic lattice dominate the response to the gravitational wave. This formulation is shown to predict a dynamical gravito-phonon Casimir effect \cite{inan2017new}, since the zero-point energy of the ionic lattice is modulated by the gravitational wave (increase in the occupation number of the lattice phonon modes), in a quantum mechanical analog of a Weber-bar effect.
%where the amplitude of the sound waves in the lattice grows, in the classical limit, due to the coupling of energy from the gravitational wave to the lattice modes.
The interaction of the gravitational wave with the phonon modes clearly breaks the spatial isotropy of the system: in this regard, the coupling with the zero-point energy of the harmonic oscillators also breaks the isotropy of the vacuum energy associated with the ionic lattice.\par
By relating the gravitational strain of space to the strain of matter, in \cite{inan2017interaction,inan2017new} the charge separation is described in terms of a different response of the ionic lattice and Cooper pair density to a gravitational wave. This difference in strain has a quantum mechanical origin, namely, the difference between the BCS energy gap (preserving the Cooper pairs) and the zero-point energy of the lattice phonon modes. Therefore, it is a macroscopic quantum effect which has no classical analog. The phenomenon could be used to quantify how gravitational waves interact with superconductors, giving rise to a situation where the former may be expelled from the interior of the latter in a gravitational Meissner-like effect.

\section{Interaction of the gravitational field with coherent macrosystems}
\label{theory}
\subsection{Generalized fields and potentials}
Schiff and Barnhill \cite{schiff1966gravitation} have shown that the electric field inside of a conductor does not vanish when a gravitational field is present. If the conductor is immersed in the weak Earth gravitational field, the quantity which vanishes in its interior is not the electric vector $\mathbf{E}$ but rather the sum %
\begin{equation}
\mathbf{E'}=\mathbf{E}+\frac{m}{e}\,\mathbf{g}\:,
\label{eq:SchiffBarnE}
\end{equation}
where $m$ and $e$ are the mass and charge of the charge carriers. A similar phenomenon holds for magnetic fields inside of superconductors. Whenever matter is in motion near a superconductor, so that a Lense-Thirring field is present, the Meissner effect requires the vanishing not of the magnetic field $\mathbf{B}$ but of the field \cite{DeWitt:1966yi}
\begin{equation}
\mathbf{G}=\mathbf{B}+m\,\mathbf{h_{0}}\:,
\end{equation}
with
\begin{equation}
\mathbf{h_{0}}=(h_{01},h_{02},h_{03})\,, \qquad
h_{ij}=g_{ij}-\eta_{ij}\:.
\end{equation}
Moreover, it is the flux of the latter combination, through any superconducting ring, which gets quantized in units of $h/2e$ rather than the magnetic field alone. Experimentally, if a superconductor is a uniform circular ring surrounding
a concentric, axially symmetric, quasirigid mass. Suppose the mass, initially
at rest, is set in motion until a constant final angular velocity is reached.
If $\mathbf{B}$ is initially zero then so is $\mathbf{G}$. Because of the flux quantization condition the flux of $\mathbf{G}$ through the superconducting
ring must remain zero. But since $\nabla\times\mathbf{h_{0}}$ is nonvanishing in the final state, a magnetic field must be induced. Then the magnetic field must arise from a current induced in the ring. The current arises from an induced motion of electrons on the surface of the superconductor.
%%%%%%%%%%%%%%%%%%%%%%%%%%%%%%%%%%%%%%%%%%%%%%%%%
\subsection{Josephson effect induced by gravity}
Ummarino and Gallerati proposed a simple theoretical model to investigate the interplay between gravitation and a superconductivity \cite{Ummarino:2020loo}.
Exploiting the above description in terms of generalized fields, they examined the possibility of a Josephson-like effect induced by the weak-static Earth's gravitational field.\par
The Josephson effect consists in the transmission of supercurrents between two superconductors separated by thin insulating barrier (thickness of the order of nanometer) by means of quantum-mechanical tunnelling of Cooper pairs \cite{Josephson:1962zz}. A simple manifestation of the Josephson effect can be observed in a circuit closed on a {superconductor-insulator-superconductor} (SIS) junction, to which a constant potential difference $\Delta V$ is applied. The voltage, in turn, produces a sinusoidal superconductive current across the junction with pulsation \cite{Barone1982physics} $\omega=\frac{2\,e\,\Delta V}{\hbar}$.
A corresponding AC supercurrent then appears in the weak link, due to the existing finite potential difference in the junction. The supercurrent density turns out to be
\begin{equation}
J_\text{s}=J_{0}\,\sin\left(\gamma_{0}+\frac{2\,e\,\Delta V}{\hbar}\,t\right)\:,
\end{equation}
where $\gamma_{0}$ is an integration constant and $J_{0}$ is the density current amplitude.\par
To study the interplay between the gravitational field and the supercondensate, generalized electric fields and potential are introduced
\begin{equation}
\mathbf{E}=\mathbf{E}_0+\frac{m}{e}\:\mathbf{E}_\text{g}\,,\qquad\quad
V=V_0+\frac{m}{e}\,V_\text{g}\:,
\end{equation}
$\mathbf{E}_0$ and $V_0$ being the standard electric field and potential.
If it is present only the Earth's static gravitational field ($\mathbf{E}_0=0$), the latter generalized field reduces to
\begin{equation}
\mathbf{E}=\frac{m}{e}\:\mathbf{E}_\text{g}=\frac{m}{e}\:\textbf{g}\:,
\end{equation}
while the corresponding potential difference reads
\begin{equation}
\Delta V=\frac{m}{e}\,\Delta V_\text{g}
    =\int_{0}^{\ell}\!dz\;\frac{m}{e}\,g=\frac{m}{e}\,g\,\ell\;,
\end{equation}
having chosen the $z$-axis along the direction of the gravitational field. The resulting induced Josephson current has then the form \cite{Ummarino:2020loo}:
\begin{equation}
I_\text{s}(t)
    =I_0\,\sin\left(\frac{2\,e\,\Delta V}{\hbar}\,t+\varphi\right)
    =I_0\,\sin(\omega\,t+\varphi)\;.
\end{equation}
Obviously, the effect disappears when the junction is rotated in a position where the normal vector to the surface is perpendicular to the gravitational field direction.\par
For a junction involving high-$T_\text{c}$ superconductors, the thickness $\ell$ of the insulating layer must satisfy $\ell\lesssim\xi$, since the coherence length $\xi$  is of the order of $10^{-9}\,\mathrm{m}$. The pulsation
\begin{equation}
\omega=\frac{2\,e\,\Delta V}{\hbar}=\frac{2\,m\,g\,\ell}{\hbar}\:,
\end{equation}
for a junction with an insulating layer of thickness $\ell\simeq1\,\mathrm{nm}$ would result in ${\omega\simeq 1.7\times10^{-4}\,\mathrm{s}^{-1}}$, determining a corresponding period for the Josephson current ${T=2\pi/\omega\simeq3.7\times10^4\,\mathrm{s}}$.
This implies that the distinctive oscillatory behaviour can be observed only in very stable junctions, since a reasonable duration of the experiment turns out to be longer than one day.

\subsection{Covariant London equations}
Continuing on this path, Hirakawa rewrote the classical London equations in a covariant form at zero Kelvin for a superconductor in a curved space-time, that is, in the presence of the gravitational field \cite{hirakawa1975superconductors}. The described situation gives rise to an alternating electric field in the sample interior. The latter field is generated when a superconducting body of mass $m$ and density $\rho$, rotating with angular speed $\omega$, is forced to vibrate with an amplitude $\delta x$.
The resulting induced charge density on the surface of the body is of the order of %
\begin{equation}
\sigma_\text{s}\,\sim\,\frac{m\,\varepsilon_{0}\,\omega^{2}}{e}\:\delta x\:.
\end{equation}
For weak fields and small velocities, this theory reduces to the first order relations obtained by Dewitt.

\subsection{Gravity-induced quantum phase}%Papini
\label{phase}
Papini studied in detail the behaviour of superconductors in the presence of a weak stationary gravitational field \cite{papini1966london,papini1967particle,papini1967detection,papini1968gravitation,
papini1969gravity,papini1970superconducting,papini1970superconductingGW,leung1971gravity}. The latter induce a peculiar quantum phase that can be explicitly characterized \cite{papini1967particle,Papini2002quantum,papini2004quantum}. Starting from an action of the form ($\hbar=c=1$)
\begin{equation}
\mathscr{S}=- m\int\!ds=-m\int\!dx^{0}\sqrt{g_{\mu\nu}\,\dot{x}^{\mu}\,\dot{x}^{\nu}}\:,
\end{equation}
where $\dot{x}^{\mu}=\tfrac{dx^{\mu}}{dx^{0}}$, the Lagrangian reads
\begin{equation}
\mathscr{L}=-m\sqrt{g_{ij}\,\dot{x}^{i}\,\dot{x}^{j}+2\,g_{i0}\,\dot{x}^{i}+g_{00}}\:,
\end{equation}
with $i,j=1,2,3$. From $\mathscr{L}$, we get the conjugate momenta
\begin{equation}
p_i=\frac{\partial\mathscr{L}}{\partial \dot{x}^{i}}
    =-m\frac{g_{i0}\,\dot{x}^{i}+g_{00}}{\sqrt{g_{jk}\,\dot{x}^{j}\,\dot{x}^{k}
        +2\,g_{j0}\,\dot{x}^{j}+g_{00}}}\:,
\end{equation}
and the Hamiltonian
\begin{equation}
\mathscr{H}=m\:\frac{g_{i0}\,\dot{x}^{i}+g_{00}}{\sqrt{g_{jk}\,\dot{x}^{j}\,\dot{x}^{k}+2\,g_{j0}\,\dot{x}^{j}+g_{00}}}\:.
\end{equation}
In the weak field approximation $g_{\mu\nu}\simeq \eta_{\mu\nu}+h_{\mu\nu}$ the Hamiltonian reads \cite{papini1967particle,Papini2002quantum,papini2004quantum}
\begin{equation}
\mathscr{H}\,\simeq\, \sqrt{p^{2}+m^{2}}\left(1+\frac{\eta_{00}}{2}\right)+\frac{1}{2}\,\frac{h^{ij}p_{i}p_{j}}{\sqrt{p^{2}+m^{2}}}-p^{i}\eta_{i0}\:.
\end{equation}
In presence of electromagnetic fields and in the low--velocity regime, this Hamiltonian leads to Schrödinger equation
\begin{equation}
i\frac{\partial\psi(x,t)}{\partial t}
=\frac{1}{2\,m}\left((p_{i}-e\,A_{i}+m\,h_{0i})^{2}-e\,A_{0}+\frac{1}{2}\,m\,h_{00}\right)\,\psi(x,t)\:.
\label{eq:Schrodinger}
\end{equation}
In normal units, the solution of this generalized Schrodinger equation reads
\begin{equation}
\psi(x,t)=\exp\left(i\:\frac{m\,c}{\hbar}\int\!h_{0i}\,dx^{i}-i\:\frac{e}{\hbar\,c}
\int\!A_{i}\,dx^{i}\right)\;\psi_{0}(x,t)\:,
\end{equation}
where $\psi_{0}(x,t)$ is the solution of the field free Schrodinger equation. If the electron-lattice interaction is added to \eqref{eq:Schrodinger}, then the resulting equation can be applied to the study of BCS superconductors in weak stationary gravitational fields \cite{papini1968gravitation,papini1969gravity,
papini1970superconducting,leung1971gravity}.
The above discussion led to some remarkable conclusions.
\begin{enumerate}[i),leftmargin=*,labelsep=2ex,itemsep=2ex]
\item The resulting relation
    \begin{equation}
    \mathbf{\nabla}\left(A_{0}-\frac{m\,c^{2}}{2\,e}\,h_{00}\right)=0
    \end{equation}
    implies that the gravitational field generates an electric field inside the superconductor, in opposition to the gravity-free case, where $h_{00}=0$. In principle, one could therefore detect a gravitational contribution by means of the electric field produced inside the sample. In the Newtonian limit, $\mathbf{E_{g}} = m\mathbf{g}/e$, which is exactly the field predicted by Schiff and Barnhill \cite{schiff1966gravitation}.
\item  Inside the superconductor one finds
    \begin{equation}
    \mathbf{B}+\frac{m\,c^{2}}{2\,e}\varepsilon_{ijk}\,\partial^{j}h^{0k}=0\:,
    \end{equation}
    where $\mathbf{B}$ is known to vanish in the absence of gravitational fields.
\item It is the total flux
    \begin{equation}
    \oint dx^{i}\,\left(A_{0}-\frac{m\,c^{2}}{2\,e}\,h_{0i}\right)
        =\frac{n\,\hbar\,c}{2\,e}
    \end{equation}
    that is quantized, rather than just the flux of $\mathbf{B}$. This again means that $h_{0i}$ could be measured, if the magnetic field it generates is sufficiently large.
\end{enumerate}
When the superconductor rotates,
\begin{equation}
h_{0i}=\left(\frac{\boldsymbol{\omega}\times \mathbf{r}}{c}\right)_{\!i}\:,
\end{equation}
and we find
\begin{equation}
\mathbf{B}=\frac{2\,m\,c}{e}\,\boldsymbol{\omega}\:,
\end{equation}
which is the London moment of rotating superconductors. Furthermore a gravitationally-induced time-independent Schiff-Barnhill field can give rise to a Hall effect in type I superconductors \cite{papini1968gravitation}.\par
Time-dependent gravitational fields induce a time-dependent electromagnetic field in superconductors and normal metals; for sinusoidal variations, the
frequency is the same for both fields. It also follows that the dielectric constants of superconductors and normal metals are affected by gravity \cite{papini1970superconducting}. Furthermore, a current can be induced in a cylindrical shell of a thin superconductor of the first type, immersed in a static external magnetic field and static scalar gravitational potential. Such current is proportional to the generalized chemical potential of the superconductor \cite{piyakis1980static,papini1984infinite}.\par
Always by keeping on to study the effect of gravitational field on condensed matter systems Papini observe that gravity can play a role in critical phenomena. Topological singularities induce ground state degeneracy and break the continuum symmetry of the vacuum. They also generate momenta oscillations about an average momentum and a positive gravitational susceptibility. Gravitational analogues of the laws of Curie and Bloch have been found for a one-dimensional model. The critical temperature for a change in phase from bound to unbound vortices has also been calculated in \cite{papini2020condensation}.
Papini deals also with the extension of the Berry phase to relativistic quantum systems represented by known wave equations. Solutions that are exact to first
order in the metric deviation $h_{\mu\nu}$ are given for Klein–Gordon, Maxwell–Proca, Dirac and spin-2 equations. The corresponding Berry phases are expressed in terms of $h_{\mu\nu}$ \cite{lambiase2021mashhoon}.

%%%%%%%%%%%%%%%%%%%%%%%%%
%C Kiefer and C Weber
\subsection{Mesoscopic quantum systems and gravity}
In \cite{Kiefer:2004hv} Kiefer and Weber discuss the different aspects of the interaction of mesoscopic quantum systems with gravitational fields. The authors consider the non--relativistic expansions of the Klein--Gordon and Dirac equations in the post--Newtonian approximation. After a short overview of classical gravitational waves, they discuss the possible interaction of gravity with quantum fluids, the latter acting as generators and/or detectors of gravitational waves in a laboratory system. Finally, they consider interaction processes involving weak gravitational fields coupled to rotating superconductors, also proposing connected experimental settings.

\subsection{Role of local vacuum energy terms in quantum gravity}
\label{role}

In \cite{Modanese:1995tx} it was proposed that the coupling of gravitation with coherent condensed matter could give an effective local contribution to the vacuum energy density (also called the ``cosmological term'' in General Relativity). This might in turn generate instabilities in the gravitational field \cite{Modanese:1996zm}. Such instabilities could be enhanced by ``dipolar fluctuations'' which are present in quantum theories of gravity based on the Einstein action \cite{modanese1999virtual,modanese2000large,modanese2000paradox}.
This can be seen as an alternative realization of the Casimir effect, which has been proven to exist also in gravitation \cite{Quach:2015qwa}.
For example, if coherent matter is described in flat space by a scalar field with action
\begin{equation}
    \mathscr{S}_\phi=\int d^4x\, \mathscr{L}_\phi=\int d^4x \left( \frac{1}{2} \partial^\mu \phi\, \partial_\mu \phi - \frac{1}{2} m^2_\phi \phi^2 + k\,\phi^4 \right)
\end{equation}
the gravitational coupling introduces a $\sqrt{g}$ volume factor. We suppose that the dynamics of $\phi$ is driven by external forces, so the coupling amounts to an external perturbation $H_\Lambda$. We call $H_\Lambda$ a local vacuum energy term generated by coherent matter and described by a macroscopic wavefunction $\phi$ that is equivalent to a classical field:
\begin{equation}
    H_\Lambda(t,\textbf{x})=\frac{1}{8\pi G} \sqrt{g(t,\textbf{x})} \Lambda(t,\textbf{x})=
    \frac{1}{8\pi G} \sqrt{g(t,\textbf{x})} L_\phi(t,\textbf{x})
\end{equation}
It is reminiscent of the photon mass term induced in QED by the coupling of the e.m.\ field to the wavefunction of the pairs in a superconductor. (A modified gauge theory of gravity which might account for anomalous effects in superconductors has been proposed by Wu \cite{Wu:2003aq}.)

The local $\Lambda$ term is much larger than the cosmological background. The currently accepted value for the cosmological background is of the order of $\Lambda c^4/G=10^{-9}$ J/m$^3$, while for instance one typically has $\Lambda c^4/G=10^6 - 10^8$ J/m$^3$ in superconductors. These values are obtained from a relativistic version of the Ginzburg-Landau Lagrangian which yields the following expression of $\Lambda$ as a function of the pairs density $\rho$ \cite{Modanese:2001wv}:
\begin{equation}
    \Lambda(t,\textbf{x})=-\frac{1}{2m} \left[ \hbar^2 (\nabla \rho)^2 + \hbar^2 \rho \nabla^2 \rho - m\beta \rho^4 \right]
\end{equation}
Here $m$ is the pair mass and $\beta$ is the second Ginzburg-Landau coefficient. Oscillations of $\rho$ in time can have typical frequencies in the MHz--GHz range, while spatial variations occur on a scale $\xi$ (the superconductor coherence length).

\subsubsection{Virtual masses, their interactions and coupling with the $\Lambda$ term}

In a paper from 2007 \cite{modanese2007vacuum} it was shown the existence in quantum gravity of a special set of vacuum fluctuations, called ``zero-modes of the Einstein action'', which are equivalent to virtual masses with a continuum spectrum. In \cite{modanese2012quantum} an approximate calculation of the pair interactions of zero-modes was given, based on the Einstein-Maxwell approximation of General Relativity and on Fermi's Golden Rule. According to this calculation, a couple of virtual masses with the same mass can form a symmetrical bound state $\Psi_S$ (ground state) and an antisymmetric bound state $\Psi_A$ (first excited state). Excitation transitions $\Psi_S \to \Psi_A$ can occur via coupling to a time-dependent vacuum energy term $\Lambda(t)$, while decays $\Psi_A \to \Psi_S$ lead to spontaneous or stimulated emission of virtual gravitons of spin 1. The energy difference $\Delta E$ between the first excited state and the ground state depends on the mass of the virtual pair; knowing the characteristic frequency of the ``pump'' $\Lambda(t)$ (typically $10^7 - 10^9$ Hz in experiments with superconductors) one can estimate the virtual masses involved at $\sim 10^{-13}$ Kg. The virtual gravitons emitted in a transition $\Psi_A \to \Psi_S$ can propagate at macroscopic distance, depending on their energy/momentum ratio, and can account in principle for some aspects of the phenomenology observed in the second Podkletnov experiment and also in the first -- supposing that in the latter the superconducting disk emits a weak, quasi-stationary force beam, as compared to the strong and brief impulse observed in the second experiment.

It can be proven that only a $\Lambda$-term can excite transitions of virtual masses efficiently, while the probability of excitation by an incoherent material source is very small. Actually, with the local strength of $\Lambda$ mentioned above for superconductors, the excitation time of the transition $\Psi_S \to \Psi_A$ is so short ($10^{-23}$ s) that the overall efficiency of the process is probably limited by the rate at which pumping energy can be supplied through the superconductor \cite{junker2008conditions}. It is known that a superconductor near equilibrium is not a good match for power transfer.

Some recent numerical results on virtual masses in quantum gravity call for an adjustment of this model. Through Metropolis -- Monte Carlo simulations of the path integral of  Einstein gravity stabilized with an $R^2$ term, it was found \cite{modanese2021quantum} that the  zero-modes analytically derived in \cite{modanese2007vacuum} are present but not widespread. Instead, the entropically prevalent virtual masses do not have a continuum spectrum but are quantized and equal to the Planck mass $M_P \simeq 10^{-8}$ Kg. They also have a short lifetime $\tau$. Their action scales as $S \sim \tau T L$, being $T$ the equivalent temperature of the Metropolis algorithm and $L$ the spatial size of the masses. It follows that pair states cannot be relevant for dynamics as it happens for the almost-stationary zero-modes, because the associated transition energies would be too large. One needs to consider the interaction of the $\Lambda$-term with collective states comprising many virtual masses. This is more complicated and will require further numerical simulations.

\subsection{Exploiting Ginzburg--Landau formalism}
\label{explGL}

In order to obtain a generalization of theoretical superconductivity that includes not only the electromagnetic field but also the gravitational one, Dinariev and Mosolov proposed a minimal generalization of the Ginzburg--Landau phenomenological equations in curved spacetime \cite{dinariev1987relativistic,pollock2014gravitational}. In the absence of electromagnetic fields, the authors found that the solution of the new time dependent Ginzburg--Landau equations is different with respect to the standard ones. The difference is related to the fact that, in the non relativistic theory, the electrons energy is usually computed with respect to the bottom of the potential hole in which they move, rather than with respect to the vacuum energy. The authors also shown that a variable non--uniform weak external gravitational field can generate electric currents in a superconducting sample.

%%%%%%%%%%%%%%%%%%%%%%%%%%%
%deMatos Tajmar

\subsubsection{Rotating reference frame}
Tate and others in their famous experiment \cite{Tate:1989zz} found that the mass of Cooper pairs $m'$ in a Niobium superconductor is not exactly twice that of an electron $(2m_{e})$. In particular, they find that the ratio is given by
\begin{equation}
\frac{m'}{2\,m_{e}}=1.000084(21)\;.
\end{equation}
A possible explanation was provided by M.\ Tajmar and C.J.\ de Matos \cite{tajmar2005extended,tajmar2006gravitomagnetic,deMatos2007possible}: applying the Ginzburg–Landau theory including frame dragging effects to a rotating superconductor, they were able to express the absolute value of the involved gravitomagnetic field, the interpreting the Cooper pair mass anomaly previously reported by Tate. Following the Ginzburg–Landau theory, they integrated the current density of Cooper pairs around a closed path, including the effects of a rotating reference frame and any gravitomagnetic fields (neglected by Tate). They obtained:
\begin{equation}
\frac{m}{e^{2}n_\text{s}}\,\oint_{\Gamma}\,\mathbf{j}\cdot\mathbf{dl}
    ~=~\frac{n\,h}{2\,e}\:-\:\int_{S_{{}_\Gamma}}\!\!\mathbf{B}\cdot d\mathbf{S}
      \:-\:\frac{m}{e}\int_{S_{{}_\Gamma}}\!\!\mathbf{B_{g}}\cdot d\mathbf{S}
      \:-\:\frac{2\,m}{e}\,\boldsymbol{\omega}\cdot\mathbf{S_{{}_\Gamma}}\:,
\end{equation}
where $\mathbf{j}$ is the density of current of Cooper pairs, $n_\text{s}$ the Cooper–electron number density, $S_{{}_\Gamma}$ the area bounded by the closed line $\Gamma$ inside the superconductor, $\omega$ the angular velocity and $\mathbf{B}=-\tfrac{2\,m}{e}\,\boldsymbol{\omega}$ the London moment. Combining this equation with the same expression without gravitomagnetic fields and calling $m^{*}$ the Cooper-pair mass measured by Tate, they found that the gravitomagnetic field $\mathbf{B_{g}}$ can be expressed as
\begin{equation}
\mathbf{B_{g}}=2\,\boldsymbol{\omega}\left(\frac{m^{*}-m}{m}\right)
    \left(1+\,\frac{1}{2\,\omega\,S_{{}_\Gamma}\,e\,n_\text{s}}\,\oint_{\Gamma}\,\mathbf{j}\cdot\mathbf{dl}\right)\:.
\end{equation}
In a superconductor of thickness much greater than the London penetration depth, the current integral in the above equations can be set to zero, as there is always a path inside the superconductor in which no current is flowing (outside the London penetration depth). The gravitomagnetic field is then just a function of the angular velocity and of the mass difference of the Cooper-pairs.
According to Tate's results, the gravitomagnetic field for a thick rotating niobium superconductor would be
\begin{equation}
\mathbf{B_{g}}=1.84\times10^{-4}\;\boldsymbol{\omega}\:,
\end{equation}
which is very large, even for small angular velocities, if compared to the gravitomagnetic field produced by the Earth (about $10^{-14}$ rad s${}^{-1}$).
The authors then discuss a possible explanation for this $B_\text{g}$ value, which is so large that it must be experimentally verified \cite{tajmar2006gravitomagnetic}.
%%%%%%%%%%%%%%%%%%%%%%%%%%%%%%%%%%%%%%%%%%%%%%%%%%%%%%%%%%%%%%%%%%%%%%%%%%%%%%%%%%%%%%%%%%%%%

\subsubsection{Dark energy contribution}
C.J.\ De Matos and C.\ Beck discuss recent laboratory experiments with rotating superconductors and show that some unexplained experimentally observed effects could be physically interpreted in terms of a possible interaction of dark energy with Cooper pairs \cite{deMatos2007possible,deMatos2008electromagnetic}. In particular they focused on anomalous acceleration signals \cite{tajmar2007measurement}, anomalous gyroscope signals \cite{tajmar2008investigation}, Cooper pair mass excess \cite{Tate:1989zz}.\par
The existence of dark energy in the universe, as indicated by numerous astrophysical observations, represents one of the most challenging problems in current theoretical physics. A great variety of different models exist for dark energy chracterization, but none of these models can be regarded as entirely convincing so far. In particular, the cosmological constant problem (smallness of the cosmologically observed vacuum energy density) remains an unsolved issue.
\par
A possible solution of this puzzle is represented by Beck and Mackey's model \cite{beck2008electromagnetic}, a Ginzburg--Landau description of electromagnetic dark energy where gravitationally active photons acquire mass in a superconductor.
This Ginzburg--Landau model is constructed to generate a cutoff for the gravitational activity of vacuum fluctuations. In this model, the authors assumed
that vacuum fluctuations of any particle can exist in two different phases: a gravitational--active one (contributing to the cosmological constant $\Lambda$) and a gravitationally--inactive one (not contributing to $\Lambda$). The model exhibits a phase transition at a critical frequency, which makes the dark energy density in the universe small and finite.\par
The above approach has many analogies with the physics of superconductors, and in particular it allows for a possible interaction between dark energy and Cooper pairs. In the Beck and Mackey's model, dark energy couples to superconducting matter only (and not to matter in the normal state): if dark energy can interact with superconducting matter only, there are no contradiction from cosmological observations, since almost all the matter in the universe is not in a superconducting state. The properties of superconductors can be then understood from the spontaneous breaking of electromagnetic gauge invariance, when the material is in the superconductive phase. In quantum field theory, this symmetry breaking leads to massive photons via the Higgs mechanism.
In this case the Maxwell equations transform to the so-called Maxwell-Proca equations, which are given by
\begin{equation}
\begin{alignedat}{2}
&\nabla\cdot\mathbf{E}=\frac{\varrho_{e}}{\varepsilon_{0}}-\frac{\phi}{\lambda^{2}_{\gamma}};
&&\nabla\cdot\mathbf{B}=0 \,,
\\[2.5\jot]
&\nabla\times\mathbf{E}=-\frac{\partial\mathbf{B}}{\partial t}\,,\qquad\qquad
&&\nabla\times\mathbf{B}=\mu_{0}\,\mathbf{j}
    +\frac{1}{c^2}\,\frac{\partial\mathbf{E}}{\partial t} -\frac{\mathbf{A}}{\lambda^{2}_{\gamma}}\:.
\end{alignedat}
\end{equation}
Here $\mathbf{E}$ is the electric field, $\mathbf{B}$ is the magnetic field, $\varepsilon_{0}$ is the vacuum electric permittivity, $\mu_{0}=(c^{2}\varepsilon_{0})^{-1}$ is the vacuum magnetic permeability, $\phi$ is the scalar electric potential,
$\mathbf{A}$ is the magnetic vector potential, $\rho_{e}$ is the Cooper pair charge density, $\mathbf{j}$ is the Cooper pair current density, $\lambda_{\gamma}=\tfrac{\hbar}{m_{\gamma}\,c}$ is the photon’s Compton wavelength which is equal to the London penetration depth \cite{debruyn2017massive}
\begin{equation}
\lambda_\textsc{l}=\sqrt{\frac{m}{\mu_{0}\,n_\text{s}\,e^{2}}}=\lambda_{\gamma}\:,
\end{equation}
where $n_\text{s}$ is the density of Cooper pair and $m_{\gamma}$ is the photon mass. Taking the curl of the above equation and neglecting the term coming from the displacement current, we get the following equation for the magnetic field:
\begin{equation}
\nabla^{2}\mathbf{B}-\frac{\mathbf{B}}{\lambda^{2}_{\gamma}}
    \:=\:\frac{1}{\lambda^{2}_\textsc{l}}\:\frac{m_{e}}{e}\:2\,\boldsymbol{\omega}\:.
\end{equation}
The solution of this equations for the one-dimensional case gives a magnetic field with an exponentially decaying term, together with a contribution proportional to the rotation frequency $\omega$. The latter terms correspond to the Meissner effect and the London moment, respectively, with \cite{tajmar2005extended}:
\begin{equation}
B=\mathcal{C}\,\exp\left(-\frac{x}{\lambda_{\gamma}}\right)-\frac{m_{e}}{e}\:2\,\omega\:,
\end{equation}
where $\mathcal{C}$ is an arbitrary constant.\par
In analogy with the electromagnetic fields produced by a Cooper pair condensate, described by a set of Maxwell-Proca equations, it is possible write
equations for the gravitational interaction in the weak field approximation. These generate gravitoelectromagnetic fields, according to a set of Einstein--Maxwell--Proca equations with a massive graviton \cite{argyris1997massive}:
\begin{equation}
\begin{alignedat}{2}
&\nabla\cdot\mathbf{E_{g}}=-\frac{\varrho_{m}}{\varepsilon_\text{g}}
    -\frac{\phi_\text{g}}{\lambda^{2}_{\text{g}}};
&&\nabla\cdot\mathbf{B_\text{g}}=0 \,,
\\[2.5\jot]
&\nabla\times\mathbf{E_\text{g}}=-\dfrac{\partial\mathbf{B_\text{g}}}{\partial t}
    \,,\qquad\qquad
&&\nabla\times\mathbf{B_\text{g}}=-\mu_\text{g}\,\mathbf{j_{m}}
    +\frac{1}{c^2}\,\frac{\partial\mathbf{E_\text{g}}}{\partial t} -\frac{\mathbf{A_\text{g}}}{\lambda^{2}_{\text{g}}}\:.
\end{alignedat}
\end{equation}
Here $\mathbf{E_{g}}$ is the gravitational field, $\mathbf{B_{g}}$ is the gravitomagnetic field, $\varepsilon_\text{g}=(4\pi G)^{-1}$ is the vacuum gravitational permittivity, $\mu_\text{g}=4\pi G/c^{2}$ is the vacuum gravitomagnetic permeability, $\phi_\text{g}$ is the scalar gravitational potential, $\mathbf{A_{g}}$ is the gravitomagnetic vector potential, $\varrho_{m}$ is the mass density of the gravitational condensate and $j_{m}$ the mass current density,  $\lambda_{\text{g}}=\tfrac{\hbar}{m_{\text{g}}\,c}$ is the graviton Compton wavelength and $m_{\text{g}}$ is the graviton mass.
The Einstein--Maxwell--Proca equations are then coupled to the electrodynamics of Cooper pairs.\par
A non--vanishing cosmological constant $\Lambda$ can be interpreted in terms of a vacuum energy density of expression
\begin{equation}
\varrho_\text{vac}=\frac{c^{4}}{8\pi G}\:\Lambda\:,
\end{equation}
\sloppy
which corresponds to a dark energy contribution with equation of state ${w=-1}$. The small astronomically observed value of the cosmological constant
$\Lambda=1.29\times10^{-52}$ $m^{-2}$ remains a deep mystery, since, at the Planck scale, the expected vacuum energy density from quantum field theories should be larger by a factor of the order of $10^{120}$, in complete contradiction with the observed value. To solve this problem, Beck and Mackey \cite{beck2008electromagnetic} suggested a model of dark energy that is based on bosonic vacuum fluctuations, the latter creating a small amount of vacuum energy density. In particular, they assumed that photons (or any other boson), with zero--point energy  $\epsilon=h\,\nu/2$, can exist in two different phases: the first is a gravitationally--active phase, where the zero--point fluctuations contribute to the cosmological constant $\Lambda$; the second is a gravitationally--inactive phase, where they do not contribute to $\Lambda$. This behaviour can be suitably described in terms of a Ginzburg-Landau formulation.\par
The above dark energy model can lead to measurable effects in superconductors, through a possible interaction between the bosons with the Cooper pairs in the sample. In superconductors, the density of superelectrons is described by the square of a macroscopic wave function as $|\Psi_\text{s}|^{2}$. Similarly, in the dark energy model, the number density of gravitationally active photons is described by the square of a different wave function, $|\Psi_\text{g}|^{2}$.
Both theories are described in terms of a Ginzburg--Landau free energy of the form
\begin{equation}
\begin{split}
F_\text{s}&=a\,\Psi_\text{s}^{2}+b\,\Psi_\text{s}^{4}\:,
\\[2ex]
F_\text{g}&=a'\,\Psi_\text{g}^{2}+b'\,\Psi_\text{g}^{4}\:.
\end{split}
\end{equation}
Both models exhibit a phase transition, the first at the critical temperature $T_\text{c}$ of the superconductor, the second at a critical frequency $\nu_\text{c}$ above which vacuum fluctuations are gravitationally inactive. The finite dark energy density, generated by gravitationally active vacuum fluctuations, leads to an accelerated expansion of the universe at large scales, as a consequence of the Friedmann equations. Furthemore, the model allows to observe the effects of dark energy on much smaller laboratory scales, where some responses  could be observed in the interior of coupled superconductors, exploiting the mutual influence of two Ginzburg--Landau potentials.\par
Additional hypotheses, with respect to the original model, consistent with the experimental observations, were then considered for rotating superconductors.
\begin{enumerate}[i),leftmargin=*,labelsep=2ex,itemsep=2ex]
\item Like normal photons with energy $\epsilon=h\,\nu$, gravitationally--active photons (graviphotons) with zero--point energy $\epsilon=h\,\nu/2$ acquire mass in a superconductor due to the spontaneous breaking of gauge invariance.
\item   The transition between the two graviphoton's phases (gravitationally--active, versus gravitationally--inactive) occurs at the critical temperature $T_\text{c}$ of the superconductor, which defines a cutoff frequency $\nu_\text{c}$ of zero--point fluctuations, specific for each superconductive material:
    $h\,\nu_\text{c}\simeq k_\textsc{b}\,T_\text{c}$.
\item   Graviphotons can form weakly bounded states with Cooper pairs, slightly increasing their mass from $m$ to $m'$. The binding energy is $\epsilon_\text{c}=\mu c^{2}$ and $m'=m+m_{\gamma}-\mu$. Since graviphotons are bound to Cooper pairs, their zero--point energies form a condensate with gravito--electrodynamic properties, similar to superconducting cavities.
\end{enumerate}
The formal attribution of a temperature $T$ to the graviphotons is done by comparing their--zero point energy with the energy of ordinary photons in a thermal bath at temperature
\begin{equation}
\frac{h\,\nu}{2}=\frac{h\,\nu}{\exp\left(\frac{h\,\nu}{k_\textsc{b}T}\right)-1}\:.
\end{equation}
This condition is equivalent to $h\,\nu=\ln\left(3\,k_\textsc{b}T\right)$.
Substituting the critical transition temperature $T_\text{c}$, specific for a given superconducting material, it is possible to find the corresponding characteristic critical frequency:
\begin{equation}
\nu_\text{c}=\ln\left(\frac{3\,k_\textsc{b}T_\text{c}}{h}\right)\:.
\end{equation}
Some preliminary, experimental evidences suggest that one might be on the right track with these type of theoretical models. The graviphotonic effect, the gravitomagnetic London moment and the non--classical inertia in rotating superconductive cavities are three different and experimentally observed effects, which can all be explained in terms of the proposed dark energy model, qualitatively and quantitatively. The theory ultimately relies on the simultaneous, spontaneous breaking of gauge invariance and general covariance principle in the interior of superconductors.\par
The above considerations imply that the dark energy of the universe produces different measurable effects, not only on cosmological scales, but also in the interior and in the vicinity of superconductors. This opens up the way for a variety of new possible laboratory experiments, testing the nature of dark energy and constraining its interaction with Cooper pairs. In this model, the gravitationally--active vacuum fluctuations underlying dark energy lead to a strong enhancement of gravitomagnetic fields, in quantitative agreement with the anomalies observed in the experiments of \cite{tajmar2007measurement,tajmar2008investigation,Tate:1989zz}.

%%%%%%%%%%%%%%%%%%%%%%%%%%%%%%%%%%%%%%%%%%%%%%%%%%%%%%%%%%%%%%%%%%%%%%%%%%%%%%%%%%%%%%
\subsubsection{Superfluidity}
The measurement of the Cooper pairs inertial mass by Tate et al.\ revealed an anomalous excess of mass. It is possible to interpret these experimental results in the framework of the discussed dark energy electromagnetic model for a superconductor physical vacuum. Being the latter associated with a preferred frame, the speed of light is variable in superconductors vacuum, giving rise to a possible breakdown of the weak equivalence principle for Cooper pairs \cite{dematos2010physical}.\par
A similar situation also occurs in the presence of superfluids. Due to the breaking of gauge symmetry in rotating superfluid Helium, the inertial mass of a vortex diverges logarithmically with the vortex size. The vortex inertial mass is therefore much higher than the classical inertial mass of the vortex core. An equal increase of the vortex gravitational mass is then questioned. The vortices in a rotating superfluid could break the weak equivalence principle in relation to a varying speed of light in this superfluid vacuum \cite{dematos2011vortices,dematos2012modified}.
%%%%%%%%%%%%%%%%%%%%%%%%

\subsubsection{Gravitational waves mirroring}
\label{subsubsec:mirror}
Other ideas and techniques were also developed, exploiting the predictions of general relativity and quantum mechanics to study peculiar situations in which gravitational radiation exhibits some kind of interplay with coherent quantum fluids. A minimal rule for coupling the electron spin to curved spacetime in general relativity implies the possibility of a related coupling between electromagnetic and gravitational radiation, mediated by the coherent superfluid.\par
Thin superconducting films are predicted to be highly reflective mirrors for gravitational waves at microwave frequencies \cite{Minter:2009fx}. The quantum mechanical non-localizability of the negatively--charged Cooper pairs, which is protected from the localizing effect of decoherence by an energy gap, causes non-geodesic motion in the presence of a gravitational wave. The latter motion is an accelerated motion through space, leading to the existence of mass and charge supercurrents inside the superconducting film. On the other hand, the decoherence--induced localizability of positively charged ions in the lattice causes a geodesic motion, as they are carried along with space in the presence of the same gravitational wave. The resulting separation of charges leads to a virtual plasma excitation within the film, enormously enhancing its interaction with the wave compared to a neutral superfluid or normal matter. The existence of strong mass supercurrents
inside a superconducting film, in the presence of a gravitational wave, implies the specular reflection of a gravitational microwave by a superconducting film of thickness much smaller than the London penetration depth of the material, in close analogy with electromagnetism.\par
Classical gravitational fields, which obey Maxwell--like equations, interact with quantum matter, which is described using the BCS and Ginzburg--Landau theories of superconductivity, as well as a collisionless plasma model \cite{Minter:2009fx}.
In the same way, starting from Maxwell--like equations for a weak gravitational field, it is possible to demonstrate the existence of a gravitational Aharonov-Bohm effect \cite{chiao2004conceptual,Roura:2021fvd}. A connection is then established between the gravitational, vectorial Aharonov-Bohm effect and the principle of local gauge invariance for non--relativistic quantum matter interacting with weak gravitational fields. The compensating vector fields, necessary by this local gauge principle, are shown to be incorporated by the DeWitt minimal coupling rule. The non--relativistic Hamiltonian for weak, time-independent fields interacting with quantum matter is then extended to time-dependent fields, and applied to problem of the interaction of radiation with macroscopically
coherent quantum systems, including the problem of gravitational radiation interacting with superconductors.\par
The Cooper pair density can be described by means of the Ginzburg--Landau free energy density embedded in curved spacetime, while the ionic lattice is modelled by quantum harmonic oscillators coupled to gravitational waves and characterized by quasi--energy eigenvalues for the phonon modes. The formulation predicts the possibility of the dynamical Casimir effect presented in Section \ref{subsec:Casimir}, the zero-point energy of the ionic lattice phonons being modulated by the gravitational wave.

\subsubsection{Gravitational field outside the sample. Fluctuations regime}
\label{sec:outside}
Inspired by the experimental effects of gravitational fields on superfluids and supercurrents, Ummarino and Gallerati considered the possible back-reaction of the latter on the surrounding local gravitational field in some specific, favourable situations \cite{Ummarino:2017bvz,Gallerati:2022pgh,Ummarino:2019cvw,Ummarino:2021vwc,Ummarino:2021tpz}.
In particular, they analysed how the local gravitational field can be affected by the presence of a supercondensate, exploiting the time-dependent Ginzburg-Landau equations \cite{GL2009Springer} in the regime of fluctuations \cite{Ummarino:2019cvw} and in a more general situation \cite{Ummarino:2021vwc,Ummarino:2021tpz}.
%The physics of low-temperature condensed matter systems is based on two fundamental notions: the low-energy long-living excitations (quasiparticles) and the mean field approximation. The BCS theory of superconductivity \cite{Bardeen:1957mv} is a classical example of the exploitation of both approaches mentioned.
Physical situations which cannot be consistently described in terms of the quasiparticle method or the mean field approximation are called \emph{fluctuations} \cite{quinn2009solid}. The regime in which the fluctuations come into play is, in general, a very narrow temperature range around the critical temperature.%
\footnote{For high temperature cuprate superconductors, organic superconductors, iron pnictides, low dimensional and amorphous superconducting systems, the situation changes radically due to the very small value of the coherence length, so that the temperature range of fluctuation is considerably larger.}
In particular, many effects on the superconducting phase occur while the system is still in the normal phase (just above the critical temperature) and originate from the appearance of the superconducting fluctuations themselves.\par
In \cite{Ummarino:2019cvw} the authors consider a superconducting sample near its critical temperature, where the thermodynamic fluctuations of the order parameter $\psi(\mathbf{x},t)$ are predominant. From a physical point of view, $\psi$ can be thought as the pseudowavefunction characterizing the motion of the center of mass of the Cooper pairs. More precisely, the sample is put at a temperature $T$ slightly greater than $T_\text{c}$, but sufficiently far from the transition point, where the system can be described in terms of linearized time-dependent Ginzburg--Landau equations being the order parameter very small \cite{schmid1966time,hurault1969nonlinear,schmid1969diamagnetic}.
From the calculated supercurrent density $\mathbf{j}_\text{s}(t)$, it is possible to extract the generalized electromagnetic fields and potentials characterizing the physical evolution of the system. The vector potential $\mathbf{A}(x,y,z,t)$ is obtained from
\begin{equation}
\mathbf{A}(x,y,z,t)=\frac{\mu_0}{4\pi}\int
    \frac{\mathbf{j}_\text{s}(t')\;\,dx'\,dy'\,dz'}{\sqrt{(x-x')^{2}+(y-y')^{2}+(z-z')^{2}}}\;,
\label{eq:A_from_js}
\end{equation}
where $t'=t-r/c$ is the retarded time, while the gravito--electric field is obtained from:
\begin{equation}
\mathbf{E}(x,y,z,t)\:=\,-\frac{\partial\mathbf{A}(x,y,z,t)}{\partial t}+\frac{m}{e}\,\mathbf{g}\:.
%    \:=\,-\frac{\mu_0}{4\pi}\dt[\js(t)]\;\mathcal{C}(x,y,z)+\frac{m}{e}\,\g\;.
\label{eq:E_from_A}
\end{equation}
The final generalized field then features two contributions: the second term is the standard constant weak Earth's gravity contribution, while the unconventional first term originates from the presence of the (non-constant) supercurrent density and can determine a local, additional contribution to the gravitational field $\mathbf{g}$ \cite{Ummarino:2019cvw}. The final result clearly depends on the superconducting sample shape and dimensions, as well as on the space point (outside the sample) where the gravitational fluctuation is measured \cite{Gallerati:2021ops,Gallerati:2020tyq}.\par
From a preliminary qualitative analysis, it is possible to show that the maximum perturbation value $\Delta$ of the local field is proportional to inverse of the coherence length, $\Delta\:\propto\:\xi(T)^{-1}$,
while it is easily demonstrated that the maximal effect occurs after a time interval $\tau_{0}\,\propto\,(T-T_\text{c})^{-1}$.
In light of the above discussion, an optimized experimental settings should involve a large high--$T_\text{c}$ superconducting sample at a temperature very close to $T_\text{c}$. The latter condition could help in extending the time range in which the effect takes place, while choosing an high--$T_\text{c}$ superconductor would determine an enhanced local alteration due to the short intrinsic coherence length. Finally, large dimensions for the sample give a larger integration range and a resulting stronger contribution.

\subsubsection{Gravitational field inside the sample}
In \cite{Ummarino:2021vwc,Ummarino:2017bvz} Ummarino and Gallerati consider the possible alteration of the local static gravitational field in the region inside the superconductor. To this end, they exploit the time-dependent Ginzburg--Landau equations for the supercondensate order parameter, looking for analytic solutions in the weak field condition in the simple case of isolated isotropic superconductor, immersed in the Earth's gravity.
%in the absence of external EM fields.
\par
%The local static gravity is treated as the gravitational component of the generalized gravito--electric field, exploiting the formal analogy discussed in the previous sections.
The dimensionless TDGL equations in a bounded, smooth and simply connected domain in $\mathbb{R}^{\textsc{n}}$ can be written as \cite{tang1995time,lin1997ginzburg}:
\begin{equation}\label{eq:dimlessTDGL}
\begin{split}
&\frac{\partial\psi}{\partial t} \,+\, i\,\phi\,\psi
    \,+\,\kappa^2\left({\left\lvert\psi\right\rvert}^2-1\right)\,\psi
    \,+\,\left(i\,\nabla+\mathbf{A}\right)^2 \psi=0 \:,
\\[2\jot]
&\nabla\times\nabla\times\mathbf{A}\,-\,\nabla\times\mathbf{B}
    =\mathbf{j}_\text{n}+\mathbf{j}_\text{s}
    =-\eta\,\left(\frac{\partial\mathbf{A}}{\partial t}+\nabla\phi\right)
    -\frac{i}{2}\left(\psi^*\nabla\psi-\psi\nabla\psi^*\right)
    -{\left\lvert\psi\right\rvert}^2\mathbf{A}\:,
\end{split}
\end{equation}
while the boundary and initial conditions in the dimensionless form read
\begin{align}
  \left.
  \begin{aligned}
  \left(i\,\nabla\psi+\mathbf{A}\,\psi\right)\cdot\mathbf{n}=0&\cr
  \nabla\times\mathbf{A}\cdot\mathbf{n}=\mathbf{B}\cdot\mathbf{n}&\cr
  \mathbf{A}\cdot\mathbf{n}=0&
  \end{aligned}
  \!\!\!\right\} \; \text{on }\partial\Omega\times(0,t)\;;
\qquad\quad
  \left.
  \begin{aligned}
  \psi(x,0)&=\psi_0(x)\cr
  \mathbf{A}(x,0)&=\mathbf{A}_0(x)
  \end{aligned}
  \!\!\!\!\right\} \; \text{on }\Omega\;.\qquad
\label{eq:dimlessboundary}
\end{align}
For a superconductor in the Earth's static gravity and in the absence of external EM fields, one can write
\begin{equation}
\mathbf{E}_0=0\,,\quad\; \mathbf{B}_0=0
\qquad\Rightarrow\qquad
\mathbf{E}=\frac{m}{e}\,\mathbf{E}_\text{g}\,,\quad\; \mathbf{B}=0\,,\qquad
\end{equation}
having also set to zero the $\mathbf{B}_\text{g}$ contribution, that is negligible in the Solar system \cite{mashhoon1989detection,ljubivcic1992proposed}.\par\smallskip
The authors tried to solve the above dimensionless TDGL in the convenient gauge choice $\phi=0$, i.e.\  the vanishing of the scalar potential. Moreover the order parameter $\psi\equiv\psi(\mathbf{x},t)$ is a complex function that can be written as
\begin{equation}
\psi=\psi_1+i\,\psi_2\:,
\end{equation}
so that one has distinct equations for the real and imaginary parts $\psi_1$ and $\psi_2$. If we restrict to a one-dimensional configuration, the equations reads
\begin{equation}
\begin{split}
\frac{\partial\psi_1}{\partial t}&=
    \frac{\partial^2\psi_1}{\partial x^2}+A\,\frac{\partial \psi_2}{\partial x}
    +\psi_2\,\frac{\partial A}{\partial x} -\psi_1\,A^2
    -\kappa^2\left({\left\lvert\psi_1\right\rvert}^2+{\left\lvert\psi_2\right\rvert}^2-1\right)\psi_1\:,
\\[2\jot]
\frac{\partial\psi_2}{\partial t}&=
    \frac{\partial^2\psi_2}{\partial x^2}-A\,\frac{\partial\psi_1}{\partial x}
    -\psi_1\,\frac{\partial A}{\partial x}-\psi_2\,A^2
    -\kappa^2\left({\left\lvert\psi_1\right\rvert}^2+{\left\lvert\psi_1\right\rvert}^2-1\right)\psi_2\:,
\\[2\jot]
\eta\,\frac{\partial A}{\partial t}&=
    -\left(\psi_2\,\frac{\partial \psi_1}{\partial x}
    -\psi_1\,\frac{\partial\psi_2}{\partial x}\right)
    -\left(\psi_1^2+\psi_2^2\right)A\:.
\end{split}
\end{equation}
Then, the authors considered an half-infinite superconducting region
%The $\mathbf{\ux}$ direction is orthogonal to the superconducting separation surface, corresponding to the $yz$ plane and parallel to the ground, so that for $x>0$ there is an empty space, while the superfluid region is located at $x\leq0$. The whole setting is
immersed in the Earth's uniform and static gravitational field. The latter is captured by the gravito--electric component $\mathbf{E}_\text{g}^{\textsc{ext}}=-g\,\vec{u}_x$, $g$ being the standard gravity acceleration, while the \emph{dimensional} form of the gravitoelectric field \emph{inside} the superfluid region is $\mathbf{E}_\text{g}=-\tfrac{\partial\mathbf{A}_\text{g}(t)}{\partial t}$. The external (\emph{outside}) gravitational vector potential is expressed as
$\mathbf{A}_\text{g}^{\!\textsc{ext}}(t)=g\left(C+t\right)\,\vec{u}_x$, $C$ being a constant, and the generalized external potential in the 1D \emph{dimensionless} form reads
$A^{\textsc{ext}}=g_\star\left(c_1+t\right)$, with $g_\star\ll 1$.\par
It is possible to express $\psi_1$, $\psi_2$ and $A$ as:
\begin{equation}
\begin{split}
\psi_1(x,t)&=\psi_{10}(x)+g_\star\,\gamma_1(x,t)\:,
\\[1.5\jot]
\psi_2(x,t)&=\psi_{20}(x)+g_\star\,\gamma_2(x,t)\:,
\\[1.5\jot]
A(x,t)&=g_\star\,\beta(x,t)\:,
\end{split}
\end{equation}
where $\psi_{10}$ and $\psi_{20}$ characterize the unperturbed system.
The sample transition to the superconducting state occurs at $t=0$ and, before the transition, no alteration of the gravitational field takes place (material in the normal state), the gravitational field assuming the same value inside and outside the sample region for $t<0$.
The final form for $\beta(x,t)$ away from borders $(\psi_{10}\simeq1,\, c_1\simeq\eta)$ turns out to be:
\begin{equation}
\beta(x,t)=e^{-\tfrac{t}{\eta}}\,\left(\eta
    +\frac{1}{\eta}\;\int^t_0\!\!dt\;e^{\tfrac{t}{\eta}}\;
    \frac{\partial\gamma_2(x,t)}{\partial x}\right)\:,
\end{equation}
from which one obtains the ratio
\begin{equation}
\frac{\mathbf{E}_\text{g}}{g_\star}=-\frac{\partial \beta(x,t)}{\partial t}=
    \frac{1}{\eta}\,e^{-\tfrac{t}{\eta}}\,\left(\eta
    +\frac{1}{\eta}\;\int^t_0\!\!dt\;e^{\tfrac{t}{\eta}}\;
    \frac{\partial\gamma_2(x,t)}{\partial x}\right)
    -\frac{1}{\eta}\frac{\partial\gamma_2(x,t)}{\partial x}\:.
\label{eq:ratio_Eg_g}
\end{equation}
The discussed formulation characterize more explicitly the proposed interplay between gravity and supercondensates in the presented, simplified setup. First, the external gravitational vector potential seems to play a role in the superconducting transition: in particular, the external constant $c_1$ tends to assume fixed values depending on the specific properties of the sample undergoing the superconducting transition. On the other hand, the back-reaction on the local gravitational to take place only after the transition itself, when the vector potential begins to ``perceive'' the presence of a superfluid phase.
The above \eqref{eq:ratio_Eg_g} for the ratio $\mathbf{E}_\text{g}/g_\star$ can be used to estimate the value of gravitational field inside the superconductor just after the superconducting phase transition:
\begin{equation}
t \simeq 0^{+}\::\qquad
    \frac{\mathbf{E}_\text{g}}{g_\star}\:\simeq\:1-\frac{t}{\eta}-\frac{1}{\eta}\frac{\partial \gamma_2(x,0^+)}{\partial x}\:.
\label{eq:Eg_g_t0}
\end{equation}
In the superconducting state, the alteration of the local field depends on physical characteristic of the involved sample. In particular, \eqref{eq:Eg_g_t0} shows that the relevant quantities are $\eta$, and the spatial derivative of $\gamma_2$.\par
In order to enhance the interaction, the variation $\tfrac{\partial\gamma_2}{\partial x}$ has to be maximized, an effect than can be achieved by introducing suitable disorder in the material sample.
A maximized effect would also require small values for $\eta$.
The latter is proportional to the product of the diffusion coefficient $\mathcal{D}$ times the conductivity just above $T_\text{c}$. This would suggest to consider materials that are bad conductors in the normal state and have low Fermi energies (for example, cuprates \cite{gonnelli2001evidence,ummarino1997breakdown}).
Performing measurements at a temperature close to $T_\text{c}$ would give rise to enhanced effects: for example, in the case of \,Bi${}_2$Sr${}_2$CaCu${}_2$O${}_8$\, \big(\,${T_\text{c}\simeq109\,\mathrm{K}}$)
there is a reduction of the local gravitational field of the order of $2\times10^{-5}$, see eq.\ \eqref{eq:Eg_g_t0} neglecting the last term.
The above analysis shows how a perceptible affection of the local field inside the sample is possible even in this simplified setup.

\subsubsection{Switching on EM fields. Vortex lattice}
A more general situation was studied by Ummarino and Gallerati in \cite{Ummarino:2021tpz}, considering a superconducting sample with finite thickness $L$ and very large dimensions along the other orthogonal directions. The sample is immersed in an external magnetic field $\mathbf{B}_0=B_0\,\vec{u}_z$ and has a square lattice of vortices, whose axes are directed along the magnetic field, and a constant external (standard) electric field $\mathbf{E}_0=E_0\,\vec{u}_x$ with the same direction of the gravitational field. The system also features a vector potential of the form $\mathbf{A}=B_0\,x\,\vec{u}_y$\:.
Given the simultaneous presence of the Earth's static gravity, the situation gives rise to a \emph{generalized} static field $\mathbf{E}_0'$ of the form
$\mathbf{E}_0'=\mathbf{E}_0+\frac{m}{e}\,\textbf{g}$ (see \eqref{eq:SchiffBarnE}) and a related scalar potential $\phi_0\:=\,-\,E_0'\,x$.\par
Working in the Coulomb gauge, it is possible to write a first-order expression for the dimensionless order parameter, satisfying linearized dimensionless TDGL equations, as \cite{Ummarino:2021tpz}:
\begin{equation}
\psi(x,y,t)=
    \sum_{n=-\infty}^{\infty}\! \left\lvert{c_n}\right\rvert\,\exp\left(i\,q\,n\left(y+\frac{E_0'}{B_0}\:t\right)\!\right)\;\:
    \exp\left(-\frac{\kappa^2}{2}\left(x-n\,x_0\right)^2
        +i\,\frac{E_0'}{\kappa}\left(x-n\,x_0\right)\right)\:,
\label{eq:psisum}
\end{equation}
in terms of the external magnetic field $B_0\lesssim B_{\textsc{c}{}_{2}}$ and material-dependent coefficients $c_n$. The solution describes the behaviour of an ordered vortex lattice moving under the influence of the external field $E_0'$.\par
In high-$T_\textrm{c}$ superconductors, the formation of a square lattice seems to be energetically favourable and the authors restrict to this case. The general $c_n$ coefficients could be then replaced by the correspondent expression for the square lattice $c_{{}_\Box}=(8\pi)^{\frac{1}{4}}\,\kappa$,\, i.e.\ a constant function of $\kappa=\lambda/\xi \gg 1$. The equations for the vector potential components read
\begin{subequations}
\begin{align}
\frac{\partial A_x(x,t)}{\partial t}\:&\simeq\:
    \frac{1}{\eta}\,\frac{\partial^2 A_x(x,t)}{\partial x^2}
    -\left(A_x(x,t)-\frac{E_0'}{\kappa}\right)\,c_{{}_\Box}^2\,\frac{\sqrt{\pi}}{\eta\,\kappa}\,\delta(x)+E_0'\:,
\label{subeq:Ax}
\\[1.25ex]
\frac{\partial A_y(x,t)}{\partial t}\:&\simeq\:
    \frac{1}{\eta}\,\frac{\partial^2 A_y(x,t)}{\partial x^2}
    -A_y(x,t)\:c_{{}_\Box}^2\,\frac{\sqrt{\pi}}{\eta\,\kappa}\,\delta(x)\:,
\label{subeq:Ay}
\\[1.25ex]
\frac{\partial A_z(x,t)}{\partial t}\:&\simeq\:
    \frac{1}{\eta}\,\frac{\partial^2 A_z(x,t)}{\partial x^2}
    -c_{{}_\Box}^2\,\frac{\sqrt{\pi}}{\eta\,\kappa}\,\delta(x)\:,
\label{subeq:Az}
\end{align}
\end{subequations}%   always put % here!
with initial conditions
\begin{equation}
A_x\left(x,\,0\right)=0\,,\qquad\quad
A_y\left(x,\,0\right)=B_0\,x\,,\qquad\quad
A_z\left(x,\,0\right)=0\,,\quad
\label{eq:initcond}
\end{equation}
and the generalized electric field $\mathbf{E}$ \emph{inside} the superfluid is given by
\begin{equation}
\mathbf{E}=-\frac{\partial\mathbf{A}}{\partial t}-\nabla\phi\:.
\label{eq:EdtAdxphi}
\end{equation}
To obtain analytical approximate solutions for the vector potential, the authors consider the effects of spatial--averaged \cite{sanders2007averaging} fields inside the supercondensate region. The analysis of these averaged effects gives rise to some interesting predictions.\par
The first effect is the emergence of a new component of the (generalized) electric field, parallel to the superconductor surface and directed along the external applied magnetic field. The value of this new contribution in dimensional units reads
\begin{equation}
E_{z}=\frac{4\pi\,B_\textsc{c}(T)\,\mathcal{D}}{\eta\,L}\:.
\end{equation}
\sloppy
For example, in the case of a sample of \,Bi${}_2$Sr${}_2$Ca${}_3$Cu${}_3$O${}_{10}$\, of thickness ${L=15\,\text{cm}}$ at a temperature ${T=102\,\text{K}}$ \big(\,${T_\textrm{c}\simeq107\,\text{K}}$,\: ${\lambda_0\simeq2.4\times10^{-7}\,\text{mt}}$,\: ${\xi_0\simeq1\,\text{nm}}$,\: ${\sigma^{-1}\simeq3.6\times10^{-6}\,\Omega\,\text{mt}}$,\: ${\mathcal{D}\simeq10^{-3}\,\text{mt}^2/\text{s}}$,\: $B_\textsc{c}(T)\simeq0.32\,\textrm{Tesla}$\: \cite{weigand2010mixed,piriou2008effect}\,\big), this would correspond to a resulting field
\begin{equation}
E_{z}=\frac{4\pi\,B_\textsc{c}(T)\,\mathcal{D}}{\eta\,L}
     =\frac{4\pi\,B_\textsc{c}(T)}{\mu_0\,\sigma\,L}
        \:\simeq\:77\,\frac{\textrm{V}}{\text{mt}}\;.\qquad\qquad
\end{equation}
The second expected effect is affection of the local gravitational field along the $x$ direction in the supercondensate region, where the generalized field can be expressed as
\begin{equation}
E_{x}=E_0'-E_0'\left(\frac{2\sqrt{2}\,\pi}{\eta}+1\right)\;\exp\left(-\frac{2\sqrt{2}\,\pi\,\kappa}{\eta\,L}\,t\right),
\end{equation}
from which it is possible to extract the predicted, temporary alteration and reduction of the local field \cite{Ummarino:2021tpz}.\par
Sample dimensions and chemical composition play a key role in maximizing the effects on the local gravitational field. Larger samples (i.e.\ larger values of $L$) and disordered material or bad metals (small $\eta$) would determine an increase of the time scales in which the effect manifests itself. Moreover, if the system is put at temperatures very close to $T_\textrm{c}$, there is an increase of the $\lambda$ parameter (as with the disorder) and related larger time scales. In the latter case, however, the effects of thermal fluctuations should also be taken into account \cite{larkin2005theory}.
In order to maximize the effect a crucial role would be played by suitable samples geometry, external electromagnetic fields of adequate frequency and appropriate characteristics of the material.

\section{Concluding remarks}
A deeper intertwining of different scientific areas has always proved to be a powerful tool for improving our understanding of many fascinating physical aspects of our universe.
By intersecting outcomes from condensed matter, general relativity, quantum mechanics and high energy physics, many developments can be found in a multidisciplinary environment, see e.g.\ \cite{Zurek:1996sj,volovik1990superfluid,Volovik:2000ua,Baeuerle:1996zz,Ruutu:1995qz,Garay:1999sk,Jacobson:1998ms,
Barcelo:2000tg,novello2002artificial,Barcelo:2005fc,Carusotto:2008ep,Mannarelli:2008jq,Boada:2010sh,Gallerati:2018dgm,Capozziello:2018mqy,Andrianopoli:2019sip,Gallerati:2021htm,zaanen2015holographic,
Franz:2018cqi,Kolobov:2021ynv,Gallerati:2021rtp,Sbitnev:2022khm,lambiase2021interaction}. The existence of an interplay between gravity and superconductivity has been investigated by many researchers in the last decades, due to the enormous conceptual implications and many possible applications. In particular, the interaction has been theoretically predicted by numerous authors, with very different approaches and techniques.\par
In this review, we have presented some of the most remarkable results in the field, covering different topics like gravity/superfluids interplay, superconducting gravitational detectors and interaction of the gravitational field with coherent macrosystems, also providing many experimental results. An intriguing field of research turns out to be the possible back-reaction exerted by the superfluid on the surrounding gravitational field: in this regard, there is still much to be done to achieve a complete understanding and characterization of the proposed effect.

\bigskip

\section*{\normalsize Acknowledgments}
\vspace{-2.5pt}This work was supported by the Open Access Publishing Fund of the Free University of Bozen-Bolzano. This work was partially supported by the MEPhI Academic Excellence Project (contract No. 02.a03.21.0005) for the contribution of prof.\ G.A.\ Ummarino.

\pagebreak

\bibliographystyle{mybibstyle}
\bibliography{bibliografia}

\end{document}